\documentclass{article}

\usepackage[preprint]{neurips_2026}

\newcommand{\cmark}{\textcolor{green!70!black}{\ding{51}}}%
\newcommand{\xmark}{\textcolor{red}{\ding{55}}}%

\usepackage{pifont}
\usepackage{longtable}  
\usepackage[utf8]{inputenc} %
\usepackage[T1]{fontenc}    %
\usepackage{hyperref}       %
\usepackage{url}            %
\usepackage{booktabs}       %
\usepackage{amsfonts}       %
\usepackage{nicefrac}       %
\usepackage{microtype}      %
\usepackage[dvipsnames]{xcolor}         %
\usepackage{amsmath, amssymb}
\usepackage{bm}
\usepackage{graphicx}
\usepackage{wrapfig}
\usepackage{algorithm}
\usepackage{algpseudocode}

\title{Live Music Diffusion Models: \\Efficient Fine-Tuning and Post-Training of\\ Interactive Diffusion Music Generators}

\author{%
  Zachary Novack$^*$ \\
  UC San Diego \\
  \And
    Stephen Brade\thanks{equal contribution, correspondence to \texttt{znovack@ucsd.edu, brade@mit.edu}} \\
  MIT \\
  \And
  Haven Kim \\
  UC San Diego \\ \And
    Hugo Flores García \\
  Adobe \\ \And
  Nithya Shikarpur \\
  MIT \\ \And
   Chinmay Talegaonkar \\
  UC San Diego \\ \And
    Suwan Kim \\
  MIT \\ \And
      Valerie K. Chen \\
  MIT \\ \And
        Julian McAuley \\
  UC San Diego \\ \And
          Taylor Berg-Kirkpatrick \\
  UC San Diego \\ \And
            Cheng-Zhi Anna Huang \\
  MIT 
}

\begin{document}

\maketitle

\begin{abstract}

Interactive streaming music generation promises the use of generative models for live performance and co-creation that is impossible with offline models.
However, SOTA models exist in the discrete-AR regime, requiring industrial levels of compute for both training and inference. In this work, we investigate whether audio diffusion models, with their wide support in the open-source community but non-streaming bidirectional nature, can be repurposed efficiently into interactive models accessible on consumer hardware.
By taking a critical look at the modern pipeline for block-wise outpainting diffusion, we identify critical inefficiencies during inference that result in strictly worse computational efficiency than their discrete-AR counterparts. We propose \textbf{Live Music Diffusion Models} (LMDMs), a simple modification of the generative diffusion process that recovers, and then outperforms, the inference complexity of the discrete Live Music Models (LMMs) through block-wise KV Caching.  Unlike LMMs, LMDMs further enable stable post-training alignment through our novel ARC-Forcing paradigm, reducing error accumulation without any explicit RL or reward models. We demonstrate the application of LMDMs in a number of creative domains, including text-conditioned generation, sketch-based music synthesis, and jamming. We finally show how LMDMs can be used as a generative instrument in a real artist-AI collaboration, utilizing LMDMs as a ``generative delay'' to transform musicians’ improvisation live for variable timbral effects while running locally on a consumer gaming laptop.

\end{abstract}

\newcommand{\zn}[1]{{\color{ForestGreen}{{[Zack: #1]}}}}
\newcommand{\ct}[1]{{\color{Yellow}{{[Chinmay: #1]}}}}
\section{Introduction}

\begin{figure}[t!]
    \centering
    \includegraphics[width=\linewidth]{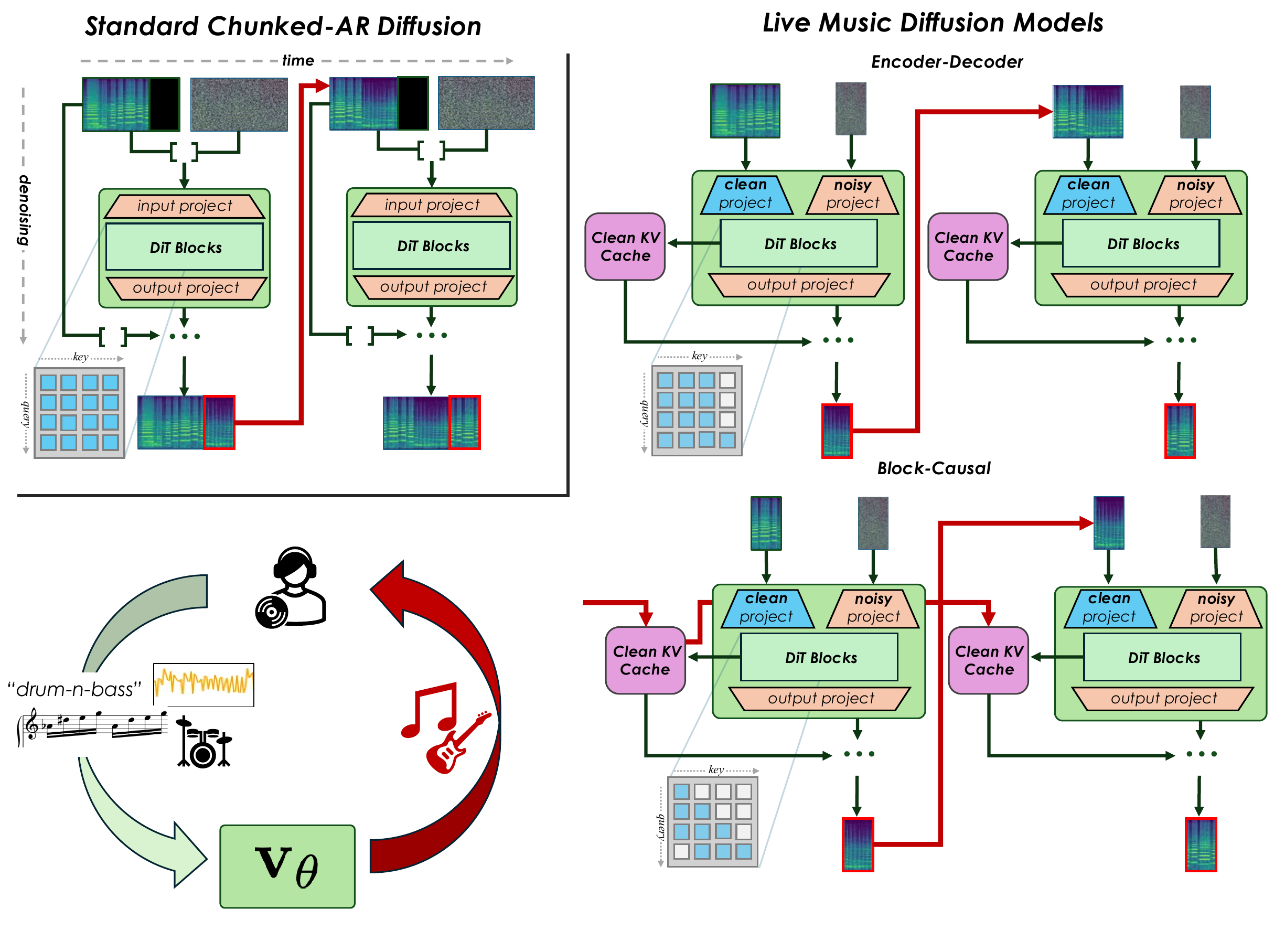}
    \vspace{-1.5em}
    \caption{\textbf{Live Music Diffusion Models}. Standard block-AR diffusion (top left) concatenates clean context with noisy states over all frames with full bidirectional attention, leaving no way to cache the context encoding despite it staying fixed for each block. LMDMs (right side) route the clean context and noisy target frames through separate projections and utilize custom attention masks to ensure clean context encoding is not impacted by the target frames, enabling KV-Caching over diffusion steps (in Enc-Dec, top-right) or both diffusion steps and or time (in Block-Causal, bottom right). LMDMs enable live interactive musical co-creation on consumer-grade hardware, modulating an array of possible controls in real time (bottom left).}
    \label{fig:main}
    \vspace{-1.5em}
\end{figure}

Generative music models have rapidly advanced, promising full song generation with high realism and control over musical attributes \citep{agostinelli2023musiclm,copet2023simple,evans2024open, Novack2025Presto, yuan2025yue}. In parallel, there has been growing interest in \emph{live interaction}: treating models as \emph{instruments} or \emph{co-musicians} to be played with in real time, 
with the recent Live Music Models (LMMs) \citep{team2025live} showing unprecedented quality while generating comprehensive musical content with live textual controls at a fixed delay. However, LMMs and other strong systems (e.g.~MusicGen-Large \citep{copet2023simple}, YuE \citep{yuan2025yue}) built on discrete autoregression (AR) have an intrinsic size bottleneck, often totaling billions of parameters: LMMs alone require over 40 GB of VRAM, making local inference on consumer hardware impractical. 

In contrast, diffusion models \citep{ho2020denoising} offer a potential solution. 
Diffusion-based approaches enjoy better data-efficiency than discrete-AR methods \citep{prabhudesai2025diffusion}, and a wealth of open-source music models exist that are performant yet much smaller than strong discrete-AR methods \citep{Novack2025Fast, evans2024open, chen2023musicldm, liu2023audioldm2}. Such methods exist in a rapidly evolving community ecosystem that has seen growing adopting by musicians in custom models and live performances \citep{foundation1, prompt_jockeys_2024, vivid_unknown}. Additionally, diffusion has shown capacity for fine-grained control, from gestural sketch conditioning \citep{garcia2025sketch2sound} to pitch, dynamics, and melodic controls \citep{tsai2025musecontrollite, Novack2024Ditto}, that have no clean analogue in discrete-AR systems. However,  diffusion-based approaches are inherently not streamable given their use of full bidirectional attention across time, and limited attempts to bridge this gap \citep{novack2025flashfoley, karchkhadze2026towards} cannot leverage the inference efficiency of discrete-AR methods (e.g.\ KV-Caching).

In this work, we repurpose open-source audio diffusion models
into interactive streaming models on consumer hardware as \textbf{Live Music Diffusion Models}\footnote{ Audio examples are available at \href{https://stephenbrade.github.io/lmdm-public/}{\texttt{https://stephenbrade.github.io/lmdm-public/}}.} (LMDMs).
By
analyzing block-diffusion outpainting, we find that a simple routing mechanism between clean history and noisy present blocks, combined with dedicated attention masking, enables noise-wise KV Caching and recovers the \emph{exact} inference complexity of encoder-decoder LMMs. A further block-causal variant achieves strictly faster complexity with full temporal KV caching. This is done \emph{purely through standard finetuning}, bypassing from-scratch training and completing in under 8 GPU hours. Second, as LMDM inference is \emph{fully differentiable} (unlike discrete-AR sampling), we 
combine the ARC framework \citep{Novack2025Fast} with Self-Forcing \citep{huang2025self} into our novel \emph{ARC-Forcing} recipe, providing global adversarial supervision on multi-block rollouts to reduce error accumulation and accelerate sampling without any RL or pretrained reward models.
Third,
we explore the full controllability of offline diffusion
across text-conditioned generation with on-the-fly prompt transitions \citep{team2025live}, localized sketch controls \citep{garcia2025sketch2sound}, and interactive accompaniment \citep{wu2025streaming}.
Finally, we demonstrate that streamability, controllability, and long-horizon stability together make LMDMs viable as generative \emph{instruments}. We build a real-time system via ONNX export and C++/JUCE, deploying sketch-conditioned LMDMs as a \emph{generative delay} on a consumer gaming laptop. We put this system in front of talented musicians from an institutional fellowship program, and are actively using LMDMs in a live musical performance.
In summary, our contributions are:
\begin{enumerate}
\item We introduce \textbf{Live Music Diffusion Models},  a simple modification to diffusion models that enables KV-Caching over diffusion steps and time through standard finetuning.
\item We propose \textbf{ARC-Forcing}, an RL-free adversarial post-training recipe providing global supervision on multi-block rollouts without reward models.
\item We bring the \textbf{full controllability} of offline diffusion, including text, sketch, and accompaniment controls, into the near real-time streaming regime.
\item We \textbf{deploy LMDMs as a generative instrument} with real musicians in collaborative sessions and live performances on consumer hardware.
\end{enumerate}

\section{Related Work}

\subsection{Interactive and Controllable Music Generation}

In the landscape of deep generative music modeling, most systems prioritize one-shot generation, mapping control modalities to fixed-length compositions. This includes high-fidelity text-to-music models \citep{evans2024open, forsgren2022riffusion, yuan2025yue} and controllable offline systems utilizing dynamics, melody, music stems, or gestural sketches \citep{Novack2024Ditto, Novack2024DITTO2DD,Wu2023MusicCM, garcia2025sketch2sound, nistal2024improving}. However, this offline paradigm remains disconnected from musical traditions centered on real-time adaptation and interaction \citep{krol_exploring_2025, kim2025amuse, agents_in_concert}, creating a workflow incompatibility for many practicing musicians. Historically, technologists and musicians have bridged gaps between technology and tradition like these by adapting expansive creative technologies to be simultaneously more accessible and more usable for musicians. For example, the miniaturization of synthesizers made them portable and affordable while adapting their control interfaces from patch cables to keyboards allowed them to be more easily integrated into musical traditions that leveraged piano.

This trajectory continues with the creation of more efficient neural architectures and inference paradigms more amenable to musical interaction. Models like RAVE \citep{caillon2021rave} which accepts audio as an input and performs real-time timbre transfer on consumer hardware exemplifies interactivity and efficiency. VampNet \citep{garcia2023vampnet} allows musicians to create generative loops, providing a generative paradigm analogous to loop pedals. Recent streaming attempts like FlashFoley \citep{novack2025flashfoley} leverage voice as a control modality to shape generated audio. The state-of-the-art model, Live Music Models (LMMs) \citep{team2025live}, brings text-controllable high-quality music generation to the near real-time setting. In this work, we push the envelope by bringing high-quality music generation to consumer-grade hardware
while simultaneously introducing controls that let musicians interface with these models through their instruments, bridging the gap between progress and tradition. To bring high-quality interactive music generation to consumer-grade hardware with the inference efficiency of discrete AR models, we introduce block-wise KV caching and an ARC-Forcing post-training paradigm inspired by the need for rollout-based stability~\citep{wu2025streaming}.

\subsection{Autoregressive Diffusion}

Many early works in the diffusion literature focused on static image generation \citep{ho2020denoising, ho2022classifier, rombach2022high, esser2024scaling}, which was the inspiration for the state of fixed-length diffusion-based music generation \citep{Evans2024LongformMG, forsgren2022riffusion, chen2023musicldm, liu2023audioldm}. Recently however, interest has grown in \emph{video} generation, and in particular autoregressive video generation, both from the lens of increasing inference efficiency \citep{yin2025slow} and for creating interactive world models \citep{genie3}. Initial diffusion-based video generation focused on approaches with bidirectional attention but folding in the noise schedule as a function of time (with future frames noisier than sooner ones), such as Diffusion Forcing \citep{chen2024diffusion} and its variants \citep{song2025history, cachay2025elucidated}. Later works expanded this to fully causal diffusion, where frames would be generated purely on a history of clean frames \citep{yin2025slow}. These have culminated with the recent \emph{Self-Forcing} paradigm \citep{huang2025self}, which post-trains fully causal video diffusion models on real rollouts from the model, using distribution matching approaches to provide exact global losses that accelerate sampling and reduce error accumulation over time. 

However, this direction remains largely unexplored for diffusion-based music generation. Recent continuous-AR approaches~\citep{pasini2024continuous, rouard2025continuous, saito2025soundreactorframelevelonlinevideotoaudio}, based on the fully autoregressive continuous language-model formulation of~\citet{li2024autoregressive}, do not study rollout-based post-training, typically require multi-billion-parameter models for strong performance, and are architecturally distinct from standard diffusion systems. Meanwhile, controllable and interactive diffusion models~\citep{novack2025flashfoley,karchkhadze2026towards} still rely on bidirectional block-wise outpainting, limiting their efficiency relative to discrete autoregressive models. In this work, we show that targeted modifications can make diffusion-based generation competitive with, and even more efficient than, the current state of the art for interactive streaming inference. We further extend Self-Forcing with Adversarial Relativistic Contrastive (ARC) post-training~\citep{Novack2025Fast} to support stable minute-long rollouts.

\section{Background}

\subsection{Flow Matching}

In this work, we primarily focus on the Flow Matching with Optimal Transport path \citep{esser2024scaling, liu2022flow} (also commonly referred as Rectified Flow) generative modeling paradigm, given its success in audio generative models \citep{Novack2025Fast, tal2025auto, lan2024high} and its general equivalence with diffusion-based approaches \citep{gao2025diffusionmeetsflow}. Given a stereo audio sequence $\mathbf{a}\in\mathbb{R}^{2 \times Lf_s}$, we first compress it into a compact, $C$-channel VAE latent representation $\mathbf{x} \in \mathbb{R}^{C \times T}$, where each $\mathbf{x}_t$ denotes the $t$th latent time frame of $\mathbf{x}$\footnote{We use $t$ to denote time, rather than noise level to be in line with existing streaming music literature \citep{wu2025streaming}.}. In flow matching, we define a forward corruption process that interpolates our sample with some amount of gaussian noise up to a noise level $k$:
$$\mathbf{x}^{(k)} = (1 - k) \cdot \mathbf{x} + k \cdot \bm\varepsilon, \quad \bm\varepsilon \sim \mathcal{N}(0, \bm{I}),$$
which we can write as shorthand as sampling $\mathbf{x}^{(k)} \sim q_k(\mathbf{x}^{(k)} \mid \mathbf{x})$, and thus $\mathbf{x}^{(0)} = \mathbf{x}$.
The goal of flow matching is to learn the reverse of this process, transferring pure gaussian noise ($k=1$) into our data distribution ($k=0$). We can view the forward process as an ordinary differential equaiton of the form:
${\mathrm{d}\mathbf{x}^{(k)}}/{\mathrm{d}{k}} = \bm\varepsilon - \mathbf{x} := \mathbf{v}$.
Thus, if we can learn a proper noise-conditioned velocity network $\mathbf{v}_\theta$ to approximate this velocity, we can solve the ODE in reverse using any normal solver (e.g. Euler, Heun, RK4). We can learn this velocity model by drawing samples from the forward corruption process and regressing our model against the marginal velocity at that point:
\begin{equation}\label{eq:flow}
\mathbb{E}_{\mathbf{x} \sim \mathcal{X}, k \sim p(k), \mathbf{x}^{(k)} \sim q_k(\mathbf{x}^{(k)} \mid \mathbf{x})}\left[\|\mathbf{v}_\theta(\mathbf{x}^{(k)}, k) - \mathbf{v}\|_2^2\right]\end{equation}
As the generative process for flow matching denotes a iterative procedure over \emph{noise levels} (from high to low) rather than any temporal axis, most flow models utilize full \emph{bidirectional} attention  across the temporal dimension, generating the entire $\mathbf{x}^{(0)}$ sequence together. We augment $\mathbf{v}_\theta$ with extra conditions $\mathbf{c}$ such as text prompts, and can sample with classifier-free guidance (CFG) as $\mathbf{v}_\theta^w(\mathbf{x}^{(k)}, k, \mathbf{c}) = \mathbf{v}_\theta(\mathbf{x}^{(k)}, k, \varnothing) - w(\mathbf{v}_\theta^w(\mathbf{x}^{(k)}, k, \mathbf{c}) - \mathbf{v}_\theta^w(\mathbf{x}^{(k)}, k, \varnothing))$ for some weight $w > 1$.

\subsection{Block-Autoregressive Outpainting}\label{sec:chard}

In this work, we broadly consider the setting of \citet{team2025live}, that is, \emph{block}-based autoregressive generation. Given some past $s$ frames of (latent) audio context, the goal is to learn a generative model over the next $o$ frames: $p_\theta(\mathbf{x}_{s:s+o} \mid \mathbf{x}_{1:s}, \mathbf{c})$. After generating the target $o$-length ``block'', the model slides its context $o$ frames in the future (forgetting the furthest history block while encoding the newly generated block as context) and continues generation. LMMs parameterize $p_\theta$ as a T5~\citep{raffel2023exploringlimitstransferlearning}-like encoder-decoder network: the encoder fuses the past history and global conditions into a single embedding that the decoder then conditions on (through cross-attention) to generate the next block, where a temporal decoder decodes autoregressively over time on the first codebook and a depth decoder decodes autoregressively over codebook levels in tandem.

Some past work has considered interactive music generation through diffusion-based block-wise outpainting \citep{karchkhadze2026towards, novack2025flashfoley}. In such setups, the flow models conditioning $\mathbf{c}$ is augmented with $s$ frames of audio context conditions 
$\mathbf{x}^{\mathrm{clean}} \in \mathbb{R}^{s \times C}$.
With most diffusion models, this is applied through \emph{channel concatenation} (i.e.~the conditions are treated as extra channels of the underlying latent $\mathbf{x}$) given the clear time-aligned nature of conditioning on clean frames. In this case, the direct input to the concatenation operation is $\mathbf{x}^{\mathrm{concat}} := [\mathbf{x}^{\mathrm{clean}}, \bm{0}_{s:T}]_{C}$ (i.e. the remaining $o = T - s$ frames are set to 0), where $[\cdot, \cdot]_{C}$ is concatenation under the \emph{channel} dimension. Fine-tuning thus proceeds nearly identically to normal flow matching: one samples $\mathbf{x}^{(k)} \sim q_k(\mathbf{x}^{(k)} \mid \mathbf{x})$, appends $\mathbf{x}^{\mathrm{clean}}$ to $\mathbf{x}^{(k)}$, and predicts the velocity. 
Once trained, inference is modified such that the generated iterate $\mathbf{x}^{(k)}$ always aligns with the ground truth $\mathbf{x}^{(0)}_{1:s}$ over the first $s$ frames, resetting these such frames to $\mathbf{x}^{(k)}_{1:s} \sim q_k(\mathbf{x}_{1:s}^{(k)} \mid \mathbf{x}^{\mathrm{clean}})$ at each step. The context window then slides (as in LMMs) over one block and inference continues, using the freshly generated block as part of the audio context. An algorithm for the inference process is given in Alg.~\ref{alg:bar}.

\begin{algorithm}
\caption{Standard Block-Wise Diffusion Outpainting}\label{alg:bar}
    \begin{algorithmic}
         \State \textbf{Input:} Model $\mathbf{v}_\theta$, Solver $\Psi$, inference steps $K$, decreasing noise level schedule $\{k_j\}_{j=1}^K$, context length $s$, target length $o$, starting reference latent $\mathbf{x}^{\mathrm{clean}}$, Number of Blocks $B$, conditions $\mathbf{c}$
         \State $\widehat{\mathbf{x}} = []$
         \For{$i = 1:B$}
            \State $\mathbf{x}^{(1)} \sim \mathcal{N}(0, \bm{I}) \in \mathbb{R}^{C \times (s+o)}$ 
            \State {\small \textcolor{purple}{\texttt{// Concatenate along time axis}}}
            \State $\mathbf{x}^{\mathrm{concat}} = [\mathbf{x}^{\mathrm{clean}}, \bm{0}_{s:s+o}]_T$ 
            \For{$j = K:1$}
                \State $\hat{\bm{v}} = \mathbf{v}_\theta([\mathbf{x}^{(k_j)}, \mathbf{x}^{\mathrm{concat}}]_C, \mathbf{c}, k_j)$
                \State $\mathbf{x}^{(k_{j-1})} =\Psi(\hat{\bm{v}},\mathbf{x}^{(k)}, k_j, k_{j-1})$
                \State {\small \textcolor{purple}{\texttt{// Reset context frames to clean latents at right noise level}}}
                \State $\mathbf{x}^{(k_{j-1})}_{1:s}  = (1 - k_{j-1}) \cdot \mathbf{x}^{\mathrm{clean}} + k_{j-1} \bm\varepsilon, \quad \bm\varepsilon \sim \mathcal{N}(0, \bm{I})$
            \EndFor
            \State {\small \textcolor{purple}{\texttt{// Concatenate target frames to output and clean latents}}}
            \State $\widehat{\mathbf{x}} = [\widehat{\mathbf{x}}, \mathbf{x}^{(0)}_{s:s+o}]_T$
            \State $\mathbf{x}^{\mathrm{clean}} = [\mathbf{x}^{\mathrm{clean}}_{o:s}, \mathbf{x}^{(0)}_{s:s+o}]_T$
        \EndFor
        \State \textbf{Return} $\widehat{\mathbf{x}}$
    \end{algorithmic}
\end{algorithm}

\section{Live Music Diffusion Models}

In this section, we show how to transform standard offline diffusion models into \textbf{Live Music Diffusion Models} (LMDMs). First, in Sec.~\ref{sec:kvcaching}, we show that a simple routing mechanism, combined with a matching attention mask, can enable both noise-wise and block-wise KV-caching. Then, in Sec.~\ref{sec:rollouts}, we show how our pipeline enables the use of RL-Free rollout adversarial post-training.

\subsection{Routing Clean Context for Efficient KV Caching}\label{sec:kvcaching}

\begin{figure}[t]
    \centering
    \includegraphics[width=\linewidth]{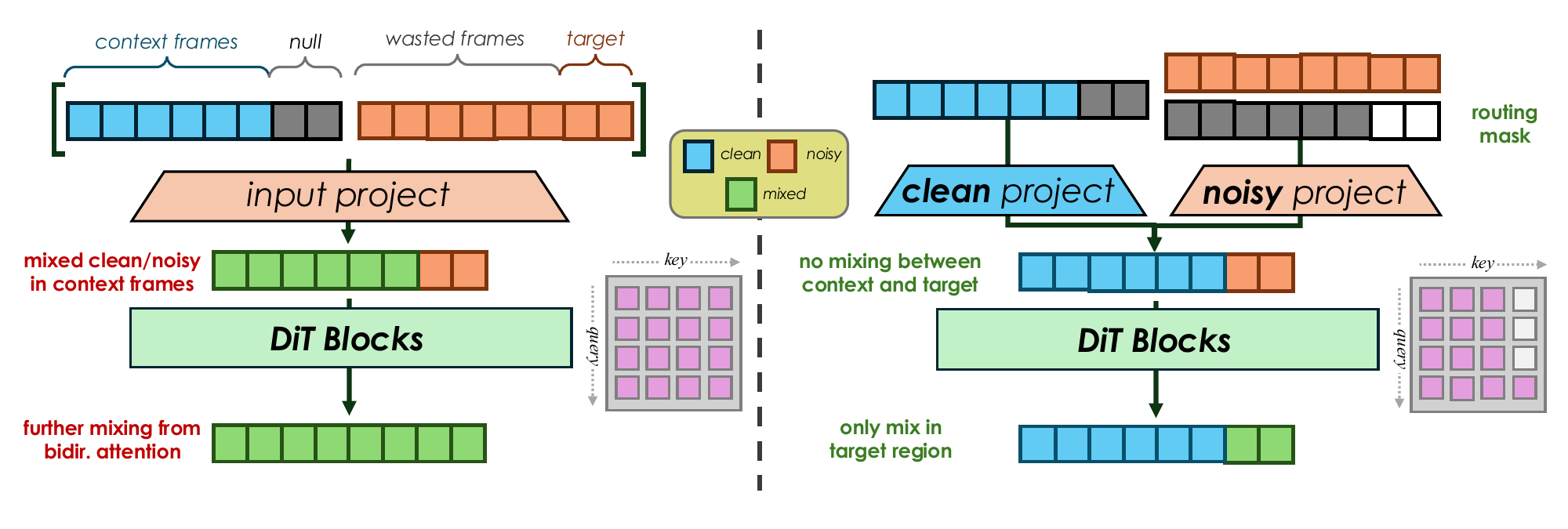}
    \vspace{-2.5em}
    \caption{Difference in initial computation graph between standard block-AR diffusion (left) and LMDMs (right, Enc-Dec version). By forcing the initial hidden state to have no mixing between clean and noisy states, and that clean frames cannot attend to noisy ones, we ensure the ability to cache clean frames for efficient inference.}
    \label{fig:routing}
\end{figure}

The key difference separating current block-wise streaming diffusion models from their LMM equivalents is the recurrent diffusion denoising process over the full latent context. First consider the inference efficiency for the encoder-decoder LMMs. If we let $\mathcal{E}_{1:s}^{\textrm{LMM}}, \mathcal{D}^{\textrm{LMM}}_t$ denote the overall latency of a \emph{single} forward pass for the encoding of $s$ frames of context and decoding of a single $t$th frame of output respectively, then the overall latency is $O(\mathcal{E}_{1:s}^{\textrm{LMM}} + \sum_{t=s}^{s+o}\mathcal{D}^{\textrm{LMM}}_t)$, where the primary bottleneck stems from the $o$ iterative decoder calls. In contrast, for normal block-based diffusion, the latency is $O((\mathcal{E}^{\textrm{Diff}} + \mathcal{D}^{\textrm{Diff}})_{1:T} \cdot K)$. Besides introducing a global dependency on the number of diffusion steps $K$ (which can be somewhat alleviated when reduced \citep{novack2025flashfoley}), this fuses the process of ``encoding/decoding'' into a joint operation over all $T$ frames that is run every diffusion step, leaving no ability to encode the context frames in a single pass as LMMs can.

The reason for this clear computational inefficiency is that standard diffusion is trained with full \emph{bidirectional} context over \emph{noisy} states, even when the model is also provided clean context through channel concatenation. Formally, for noisy states $\mathbf{x}^{(k)}$ and clean context $\mathbf{x}^{\mathrm{clean}}$, we can write the \emph{initial} hidden state of the DiT (i.e. before any attention blocks) under channel concatenation as:
\begin{equation}
    \mathbf{h}^{\mathrm{init},k} = \mathbf{W}^{\mathrm{init}} [\mathbf{x}^{(k)}, \mathbf{x}^{\mathrm{concat}}]_C = \mathbf{A}\mathbf{x}^{(k)} + \mathbf{B}\mathbf{x}^{\mathrm{concat}},
\end{equation}
where $\mathbf{W}^{\mathrm{init}} \in \mathbb{R}^{H \times 2C} = [\mathbf{A}, \mathbf{B}]_C$ is the input projection of the model and $\mathbf{A}, \mathbf{B}$ are the weight components for the noisy and clean latents respectively. However, we know from Sec.~\ref{sec:chard} that $\mathbf{x}^{\mathrm{concat}}$ is all 0s after the first $s$ context frames, thus giving us that:
\begin{equation}
    \mathbf{h}^{\mathrm{init},k}_{s:T} = \mathbf{A}\mathbf{x}^{(k)}_{s:T} \qquad \mathbf{h}^{\mathrm{init},k}_{1:s} = \mathbf{A}\mathbf{x}^{(k)}_{1:s} + \mathbf{B} \mathbf{x}^{\mathrm{clean}}
\end{equation}
This exposes one key problem: the part of our initial hidden state that ``encodes'' past context is mixed with variable noise levels, making the input to the transformer blocks change at every sampling step.
To alleviate this fact, we propose a simple solution: First, as we already know a priori which frames are context vs. target, we can implement a simple routing mask $\mathbf{r} := [\bm{0}_{1:s}, \bm{1}_{s:s+o}]_T$ (i.e.~a mask separating context from generation), which will be combined with our noisy latent before projecting into the model. This then gives us:
\begin{align}
     \mathbf{h}^{\mathrm{init},k} &= \mathbf{W}^{\mathrm{init}} [\mathbf{r} \odot\mathbf{x}^{(k)}, \mathbf{x}^{\mathrm{concat}}]_C\\
     \implies\mathbf{h}^{\mathrm{init},k}_{s:T} &= \mathbf{A}\mathbf{x}^{(k)}_{s:T} \qquad \mathbf{h}^{\mathrm{init},k}_{1:s} =\mathbf{B} \mathbf{x}^{\mathrm{clean}},
\end{align}
and thus $\mathbf{h}^{\mathrm{init},k}_{1:s}$ is the same for all possible noise levels $k$ and is only a function of the context (see Fig.~\ref{fig:routing} for a graphical demonstration).

While this guarantees that our \emph{input} to the DiT is independent of the noise level, it does not stop
our ``encoding'' representations from attending to the future, thus allowing the encoding to change through each DiT block as a function of the target block. To fix this and allow for closing the efficiency gap with LMMs, we propose two attention mask variants for LMDMs:

    \noindent \textbf{Encoder-Decoder LMDMs:} Here, we restrict every attention operation inside the DiT such that the first $s$ frames can only attend to each other and \emph{not} the last $o$ frames (i.e. where we aim to generate the next block of audio latents), while the output last $o$ frames can attend to themselves and all proceeding frames. This asymmetric attention pattern 
    fully decouples the encoding of past context from the decoding of the next block. Because of this, we can now use KV-Caching \citep{pope2023efficiently} over successive diffusion sampling steps: given a block of clean context $\mathbf{x}^{\mathrm{clean}}$, we can first pass this through our DiT and cache Key/Value states for each transformer block, and then perform every step of diffusion denoising for the target block using these states without recomputation. This yields an inference complexity of $O(\mathcal{E}^{\textrm{LMDM}}_{1:s} + \mathcal{D}^{\textrm{LMDM}}_{s:T} \cdot K)$, achieving the same complexity class as LMMs, i.e.~a \emph{single} encoding pass over clean context and an iterative decoding process for the next block. We term this as ``Encoder-Decoder'' LMDMs as it follows much of the same process of classical Encoder-Decoder LLMs (and LMMs), where the explicit encoded representation produced by a separate encoder module is replaced by an implicit encoding through the KV-Cache.
    
    \noindent \textbf{Block-Causal LMDMs:} While Enc-Dec LMDMs enable KV-Caching as a function of \emph{diffusion step}, there is no temporal KV-Caching possible, as with a fixed $s$-frame window with bidirectional attention the context changes every time we finish generating a new block and add it to the context. To enable KV-Caching over noise level \emph{and time}, we can modify the attention mask further: by introducing a block-causal dependency over $o$-sized blocks  within the first $s$ frames, we enforce that frames of context can only attend to past context (or within their block). Thus, after generating the newest block, only the newly generated block must be cached before proceeding with the next generated block (as no other context blocks attend forwards). After a warmup period to encode each $o$-sized block of the context $s$ frames, this exposes the inference complexity of $O(\mathcal{E}^{\textrm{LMDM}}_{s-o:s} + \mathcal{D}^{\textrm{LMDM}}_{s:T} \cdot K)$. Here we achieve a strictly better complexity than LMMs by removing the need to encode the whole context for each new block generation.

\begin{minipage}[t]{0.5\linewidth}
\vspace{0pt}  
    \begin{algorithm}[H]
\caption{Encoder-Decoder LMDM Inference}\label{alg:ed}
    \begin{algorithmic}     
         \State $\widehat{\mathbf{x}} = []$
         \For{$i = 1:B$}
         \State {\small \textcolor{blue}{\texttt{// Instantiate Noise for target}}}
            \State $\mathbf{x}^{(1)} \sim \mathcal{N}(0, \bm{I}) \in \textcolor{blue}{\mathbb{R}^{C \times o}}$ 
            \State {\small \textcolor{blue}{\texttt{// Build KV cache over clean frames}}}
            \State \textcolor{blue}{$\mathbf{KV} = \mathbf{v}_\theta^{\mathrm{KV}}( \mathbf{x}^{\mathrm{clean}}, \mathbf{c}, 0)$}
           \State {\small \textcolor{blue}{\texttt{// Inference only over target frames}}}
            \For{$j = K:1$}
     
                \State $\hat{\bm{v}} = \textcolor{blue}{\mathbf{v}_\theta(\mathbf{x}^{(k_j)}, \mathbf{c}, k_j \mid \mathbf{KV})}$
                \State $\mathbf{x}^{(k_{j-1})} =\Psi(\hat{\bm{v}},\mathbf{x}^{(k_j)}, k_j, k_{j-1})$
            \EndFor
            \State $\widehat{\mathbf{x}} = [\widehat{\mathbf{x}}, \textcolor{blue}{\mathbf{x}^{(0)}}]_T$
            \State $\mathbf{x}^{\mathrm{clean}} = [\mathbf{x}^{\mathrm{clean}}_{o:s}, \textcolor{blue}{\mathbf{x}^{(0)}}]_T$
        \EndFor
        \State \textbf{Return} $\widehat{\mathbf{x}}$
    \end{algorithmic}
\end{algorithm}
\end{minipage}
\begin{minipage}[t]{0.5\linewidth}
    \vspace{0pt}  
    \begin{algorithm}[H]
\caption{Block-Causal LMDM Inference}\label{alg:bc}
    \begin{algorithmic}
         
         \State $\widehat{\mathbf{x}} = []$
          \State {\small \textcolor{blue}{\texttt{// Prefill clean blocks into KV Cache}}}
         \State \textcolor{blue}{$\mathbf{KV} = []$}
         \For{\textcolor{blue}{$b = 1: \lfloor{s/o}\rfloor$}}
            \State \textcolor{blue}{$\mathbf{kv}_b =  \mathbf{v}_\theta^{\mathrm{KV}}( \mathbf{x}^{\mathrm{clean}}_{ o\cdot(b-1): o \cdot b}, \mathbf{c}, 0 \mid \mathbf{KV})$}
            \State \textcolor{blue}{$\mathbf{KV} = [\mathbf{KV}, \mathbf{kv}_b]_T$}
         \EndFor
         \For{$i = 1:B$}
         \State {\small \textcolor{blue}{\texttt{// Instantiate Noise for target}}}
            \State $\mathbf{x}^{(1)} \sim \mathcal{N}(0, \bm{I}) \in \textcolor{blue}{\mathbb{R}^{C \times o}}$ 
           \State {\small \textcolor{blue}{\texttt{// Inference only over target frames}}}
            \For{$j = K:1$}
                \State $\hat{\bm{v}} = \textcolor{blue}{\mathbf{v}_\theta(\mathbf{x}^{(k_j)}, \mathbf{c}, k_j \mid \mathbf{KV})}$
                \State $\mathbf{x}^{(k_{j-1})} =\Psi(\hat{\bm{v}},\mathbf{x}^{(k_j)}, k_j, k_{j-1})$
            \EndFor
            \State $\widehat{\mathbf{x}} = [\widehat{\mathbf{x}}, \textcolor{blue}{\mathbf{x}^{(0)}}]_T$
            \State {\small \textcolor{blue}{\texttt{// Update Cache with new block}}}
            \State \textcolor{blue}{$\mathbf{kv}_i =  \mathbf{v}_\theta^{\mathrm{KV}}( \mathbf{x}^{(0)}, \mathbf{c}, 0 \mid \mathbf{KV})$}
            \State \textcolor{blue}{$\mathbf{KV} = [\mathbf{KV}_{o:s}, \mathbf{kv}_i]_T$}
        \EndFor
        \State \textbf{Return} $\widehat{\mathbf{x}}$
    \end{algorithmic}
\end{algorithm}
\end{minipage}

\subsubsection{Efficient Finetuning and Inference}

We display the main architectural differences and attention masks in Figs.~\ref{fig:main} and ~\ref{fig:routing}. Because the modifications only change the initial projection through an element-wise mask and the underlying attention pattern, training can proceed with the standard flow matching pipeline\footnote{Note that we find extra added stability by masking the L2-based flow loss to just the target frames.} from Eq.~\ref{eq:flow}. Additionally, since the only unique parameters for embedding past context are the weights of the $\mathbf{B}$ matrix for initial state injection,  turning normal diffusion models into LMDMs can be done easily with no aggressive changes to the overall architecture. These points combined allow for an initial LMDM training phase that easily slots into modern diffusion codebases \emph{and} can work from pre-initialized diffusion checkpoints with limited overhead.

While during training we need the routing mechanism to split the input representation to their correct projection modules, for inference through KV-Caching we can avoid this altogether (see Algs.~\ref{alg:ed} and~\ref{alg:bc} for the modified inference algorithms) and accelerate performance. By separating out the forward passes for encoding context (denoted as $\mathbf{v}_\theta^{\mathrm{KV}}$) from the main sampling of the target chunk and ensuring that each each encoding pass happens in a bidirectional attention region (the full context for Enc-Dec, each context block for Block-Causal), we can use the highly optimized \texttt{flash-attention} kernels for inference, removing the need for custom attention masks with unoptimized implementations at test time.\footnote{Writing custom kernels to fuse such operations is a keen direction for future work.} This, when combined with \texttt{torch.compile}, enables fast inference with a round trip latency of 110-170ms (measured on a 6000 Pro Blackwell GPU) before our post-training step in the next section, with the KV-caching giving an approximate 20-25\% speedup per forward pass.

\subsection{Rollout Post-training through ARC-Forcing}\label{sec:rollouts}

Though the above routing mechanism and attention pattern enable block-wise AR diffusion with efficient KV-Caching, they do not fix one key issue: \emph{error accumulation}, which is a known issue in discrete-AR \citep{wu2025streaming} and continuous AR models \citep{pasini2024continuous, saito2025soundreactorframelevelonlinevideotoaudio, huang2025self}. As training only supervises the model on a single block level, it does not match the inference scenario where errors may compound as we successively condition the model on its own outputs. This thus motivates the need for an explicit \emph{post-training} phase, where one can potentially supervise the model on its own rollouts. However, such rollout post-training phases in AR music generation have both required brittle RL-based pipelines \citep{wu2025adaptive, wu2025generative} and explicitly trained reward models \citep{wu2025streaming} to do so.

In this work, we show that we can avoid both such issues. First, as diffusion sampling is itself a fully differentiable operation (in contrast to discrete-AR regimes, where inference involves sampling from the output codebook probabilities), we can avoid any need for explicit RL to post-train our model and instead rely on the recent \emph{Self-Forcing} paradigm \citep{huang2025self} from video generation. Then, to avoid the need of an explicit reward model, we can instead purely supervise the post-training process as a fully adversarial one, drawing from the recent Adversarial Relativistic Contrastive (ARC) method \citep{Novack2025Fast} and functionally learn our reward model as we post-train our generative model.

We denote the adaptation of these two approaches to streaming music generation as \textbf{ARC-Forcing}. Formally, we aim to post-train our flow model $\mathbf{v}_\theta$ into a few-step generator $G_\phi$, thus increasing inference speed, but \emph{also} to improve full AR rollouts with global supervision. To do so, we follow \citet{huang2025self} by generating $B$-block rollouts from $G_\phi$, using KV-Caching (either noise-wise or both noise and time-wise) to maintain efficient training.\footnote{While ARC-Forcing is theoretically possible for standard block-wise diffusion outpainting, the memory bandwidth quickly balloons as a functions of $B$ leading to computational intractable training.} To further accelerate post-training, we use a stochastically chosen $k \sim U[2, K_{\text{max}}]$ steps for each generated block, only propagating gradients on the final step and disabling them for previous steps and context encoding. For each batch item, $G_\phi$ receives either $s$ frames of existing context (mirroring scenarios during long-form rollouts), a null context with probability $p_{\text{uncond}}$ (mimicking the start of generation), or a partially null context where only the first uniformly sampled $n$ frames are null with $p_{\text{partial}}$.

Instead of using an explicit reward model (which is not readily available) or using the original model itself as a ``teacher'' (which requires the teacher itself be of high enough quality), we instead use a noise-aware discriminator $D_\psi$ with \emph{full bidirectional} context as our source of global supervision. $D_\psi$ is initialized from our base diffusion model (i.e.~not the finetuned LMDM), and thus can take in text conditions. Given our generated rollout $\widehat{\mathbf{x}}$ and ground truth music (with matching controls and starting context) ${\mathbf{x}}$, we add noise to both at the same level and pass both into the discriminator. We then use the \emph{relativistic} loss $\mathcal{L}_R$:
\begin{equation}
     \mathbb{E}_{\mathbf{x}, \mathbf{c} \sim \mathcal{X}, \widehat{\mathbf{x}} \sim G_\phi(\mathbf{x}_{1:s}, \mathbf{c}), k \sim p(k)}\left[f\left(D_\psi(q_k(\widehat{\mathbf{x}}^{(k)} \mid \widehat{\mathbf{x}}), \mathbf{c}, k) - D_\psi(q_k({\mathbf{x}}^{(k)} \mid {\mathbf{x}}), \mathbf{c}, k)\right)\right],
\end{equation}
where $f(x)=\log(1+\exp(x))$ is the softplus function. By supervising the model on long music-rollout pairs, we not only avoid degenerate solutions common in other GAN objectives \citep{huang2024gan} but also provide learning signal on the full rollout rather than single blocks. In order to ensure that $D_\psi$ does not overfit to high-frequency features and can encourage adherence to the underlying text prompts, we train $D_\psi$ with an auxiliary contrastive objective $\mathcal{L}_C$, using a relativistic objective on real music with the correct vs. incorrect \emph{prompts}:
\begin{equation}
     \mathbb{E}_{\mathbf{x}, \mathbf{c} \sim \mathcal{X}, k \sim p(k)}\left[f\left(D_\psi(q_k({\mathbf{x}}^{(k)}  \mid {\mathbf{x}}), \mathcal{P}(\mathbf{c}), k) - D_\psi(q_k({\mathbf{x}}^{(k)} \mid {\mathbf{x}}), \mathbf{c}, k)\right)\right],
\end{equation}
where $\mathcal{P}$ is a random batch-wise permutation matching up each music sample with a random prompt. The discriminator is then trained with a combination of $\mathcal{L}_C$ and $\mathcal{L}_R$, with a weight $\lambda$ determining the strength of the contrastive term (in our work, set to 1). Unlike in previous works \citep{Novack2025Fast, novack2025flashfoley} that used ARC in the offline setting, we found one critical change needed for adapting it to rollout-based long-form post-training: as the context window for $D_\psi$ is decoupled and is considerably larger than the context window for $G_\phi$ (i.e. $\approx30$s for $D_\psi$ vs. $\approx10$ for $G_\phi$), initializing $D_\psi$ directly from the LMDM $\mathbf{v}_\theta$ resulted in a weak discriminator that quickly destabilized post-training. To remedy this, we found that warms-tarting the backbone diffusion model of the discriminator $D_\psi$ on longer audio segments using Eq.\ref{eq:flow} for a few thousand iterations stabilized long rollout post-training. After performing ARC-Forcing, the model can now stably sample in [1, 8] steps without CFG (using the ``ping-pong'' sampler from \citet{song2023consistency}), bringing the total latency into the $\approx$30ms regime.

\begin{figure}
    \centering
    \includegraphics[width=\linewidth]{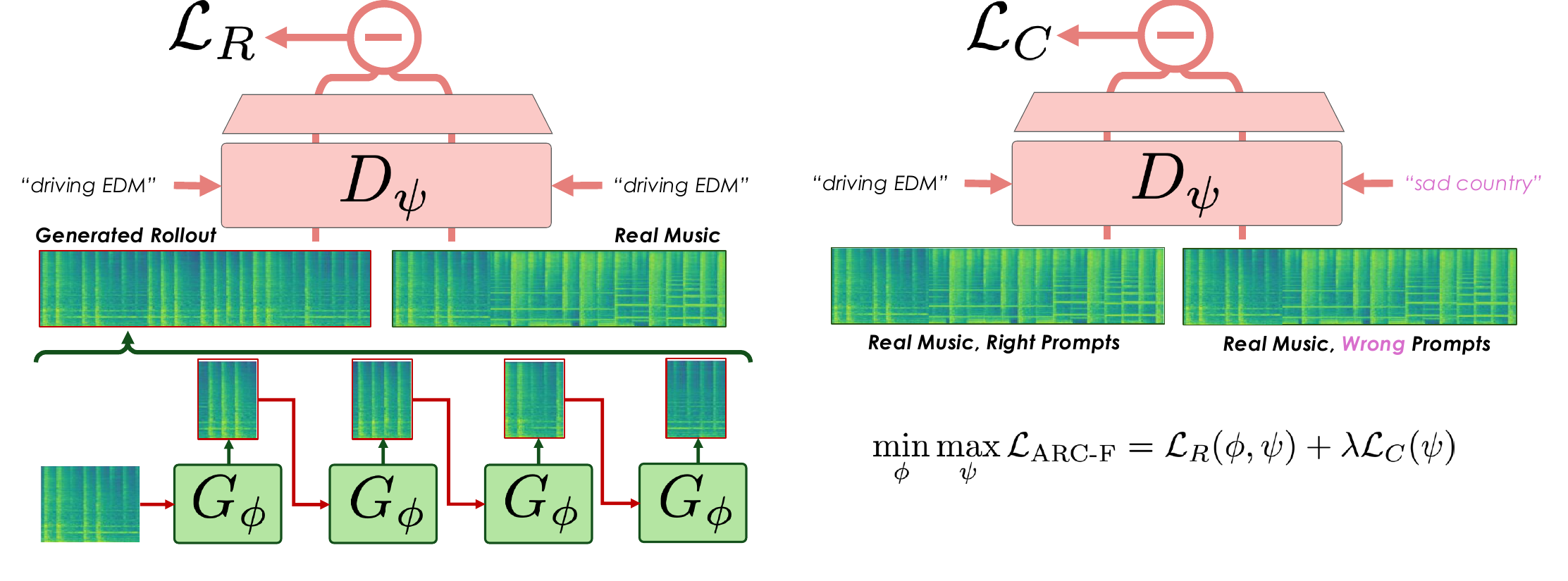}
    \vspace{-2.5em}
    \caption{ARC-Forcing. $G_\phi$ is post-trained by generating AR rollouts across time with KV Caching, and then passing in the rollouts along with real music (with the same starting context and conditions) into the bidirectional $D\psi$, which uses a relativistic objective. $D_\psi$ is also trained with an auxiliary contrastive loss on real music with matching vs. mismatched captions to encourage text following.}
    \label{fig:arcforcing}
    \vspace{-1em}
\end{figure}

\section{The Live Music Design Space}

Despite the growing interest in live interactive music models controlled with text \citep{team2025live}, sketch based controls for pitch and other features \citep{novack2025flashfoley}, or audio models that can jam \citep{wu2025streaming}, each of these works has investigated bespoke architectures in isolation between one another. Inspired by \citep{kim2026design}, in this section we unify these broad tasks and show that the all can be realized under the flexible design framework of LMDMs. Notably, as we wish to model $p_\theta(\mathbf{x}_{s:s+o} \mid \mathbf{x}_{1:s}, \mathbf{c})$, there are no restrictions on the form that $\mathbf{c}$ can take. 
We can delineate conditioning signals for these models across two axes: (1) whether controls are global ($\mathbf{c} \in \mathbb{R}$) or local ($\mathbf{c} \in \mathbb{R}^T$), and (2) whether they are \emph{instrument}-like or \emph{accompaniment}-like. In the former case, controls describe \emph{features} of the target output (e.g.~the volume curve of the generated chunk), and interaction is determined purely by how quickly the model can synthesize the current block of controls. In the latter case, controls describe \emph{conditions} that provide context to the model but arrive at a strict external time schedule decoupled from model inference(e.g.~another stream of music), and thus interaction must balance reactivity to the controls with practical latency. We hence can define many previous disparate interactive paradigms within this framework:

\noindent \textbf{Global Text-Conditioning.}
In the case which is most comparable to \cite{team2025live}, $\mathbf{c}$ is simply a global text prompt with no temporal access. Though inherently non-temporal, we classify text prompts as {instrument}-like due to the use of \emph{prompt transitions}, where such global conditions are modulated between different prompts in real time.

\noindent \textbf{Instrument-like Sketch Controls.} In the case most similar to \citep{novack2025flashfoley}, 
$\mathbf{c} \in \mathbb{R}^T$ are instrument-like
local conditions which gives the model detailed local information about what must be generated in future. Specifically, we explore top-k CQT conditions (similar to \cite{tsai2025musecontrollite}) and a loudness condition which mirrors \cite{garcia2025sketch2sound} and \cite{novack2025flashfoley}.

\noindent \textbf{Accompanying Stem Generation.} This case explores when the condition is a separate accompanying audio stream, such as the live jamming setting studied in \citet{wu2025streaming}. 
Unlike the sketch and text-conditioned settings above, the accompaniment task is purely \emph{accompaniment-like}: the model has no explicit time-aligned features for what the future block should sound like (i.e. no \emph{features} of the target block), and must instead infer a musically compatible continuation from the causal history of the single input stem and its own prior outputs, while also dealing with a fixed future visibility $t_f < 0$ (i.e. the model only receives signal from the other stem up to some cutoff frame before the target block) to compensate for system latency.

\section{Experiments and Results}
\label{sec:experiments}

\begin{table}[h!]
    \centering
    \footnotesize
    \begin{tabular}{lccccc|ccc}
    \toprule
         \textbf{Method}&  \textbf{D-NFE}&\textbf{Blocks}&  \textbf{Sampler} &\textbf{TTFF}$\downarrow$&\textbf{w/Priming?}&  \textbf{FD}$\downarrow$&  \textbf{KD}$\downarrow$& \textbf{CLAP}$\uparrow$\\
         \midrule
         Magenta RealTime&  $800^\dagger$&24&  - &$\approx4$&\xmark &  72.14&  \textbf{0.47}& 0.35\\
         Stable Audio Open&  100 &1&  DPM++ &10.35&\xmark &  96.51&  0.55& \textbf{0.41}\\
         MusicGen-Large&  2.4K&1&  - &10.81&\xmark &  190.47&  0.52& 0.31\\
         \midrule
         LMDM (ED)&  50 &21&  Euler &0.11&\xmark &  61.06&  1.14& 0.20\\
         LMDM (ED)+AF&  8 &21&  Ping-Pong &0.03&\xmark &  \textbf{35.88}&  0.74& 0.29\\
         LMDM (BC)&  50 &21&  Euler &0.17$^\ddagger$&\xmark &  64.87&  1.20& 0.20\\
         LMDM (BC)+AF&  2 &21&  Ping-Pong &0.02&\xmark &  47.26&  0.91& 0.23\\
         \midrule
         \midrule
          LMDM (ED)&  50 &21&  Euler &0.11&\cmark&  35.35&  0.62& 0.23\\
         LMDM (ED)+AF&  8 &21&  Ping-Pong &0.03&\cmark&  \textbf{29.00}&  \textbf{0.35}& \textbf{0.32}\\
         LMDM (BC)&  50 &21&  Euler &0.17&\cmark&  47.13&  0.74& 0.24\\
         LMDM (BC)+AF&  2 &21&  Ping-Pong &0.02&\cmark&  35.45&  0.53

& 0.23\\
         \bottomrule
    \end{tabular}
    \caption{Global Results on Text-Conditioned LMDMs. $^\dagger$Magenta-RTs NFE's can be broken down to 50 calls for the temporal transformer and $15 \cdot 50$ calls of the lightweight depth transformer. $^\ddagger$Despite the more efficient inference in the BC case, we empirically found their wall clock time to be slightly slower than the ED case, likely due to our suboptimal implementation.}
    \label{tab:global}
\end{table}

In this section, we evaluate our LMDMs on an array of creative musical tasks. First, we compare against existing text-conditioned generation systems in both standard generation and prompt transitions, as well investigate the roll ARC-Forcing plays in stabilizing long-form generation. Then we investigate how LMDMs perform for accompaniment generation as a function of the future visibility. Finally, we assess LMDMs capacity for sketch-conditioned generation through both offline evaluation and a musician interactive user study. Full details can be found in App.~\ref{app:evals}.

\subsection{Text-Conditioned LMDMs}

\noindent \textbf{Global evaluation.} 
Here, following \citet{team2025live} we report standard global musical metrics (FD/KL for quality, CLAP score for text adherence), as well as metrics for latency (D-NFE, TTFF). We report results both with and without ground truth audio priming. 
In Tab.~\ref{tab:global} we find that despite having half the parameters and trained using nearly 100x less data than LMMs, LMDMs show competitive quality and text adherence with drastically faster inference. We further find that in Enc-Dec LMDMs generally outperform Block-Causal ones.\footnote{We report best performance for LMDMs across NFEs in [1,8], see App.~\ref{app:evals} for more details.} This suggests that the ability to adapt the full contextual encoding as new music is added to the context, despite the fact that this may cause quicker autoregressive drift, is beneficial for performance.

\begin{figure}
    \centering
    \includegraphics[width=0.9\linewidth]{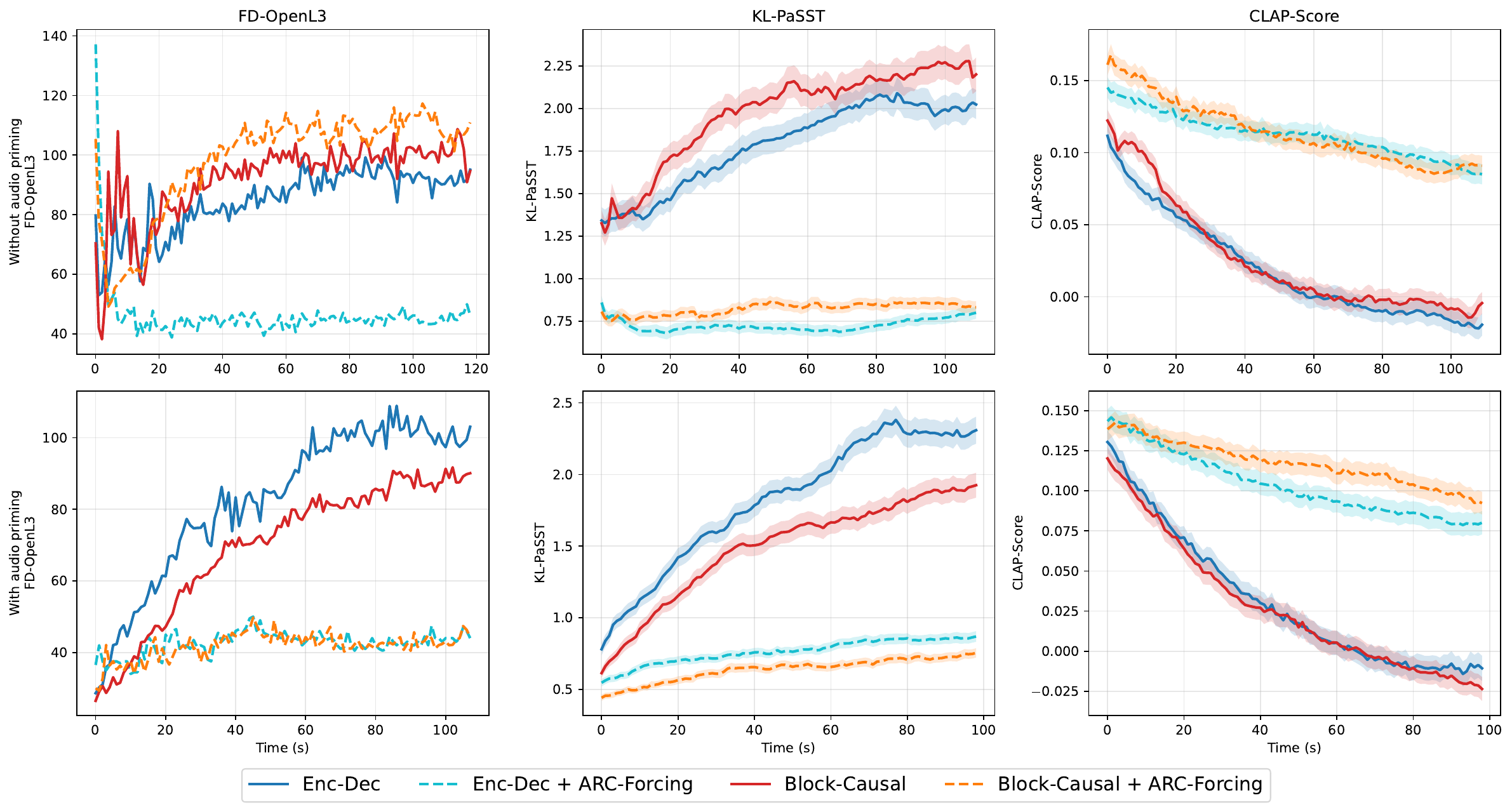}
    \caption{Global Text-Conditioned metrics over time. In both Enc-Dec and Block-Causal LMDMs, ARC-Forcing strongly reduces error accumulation and the gradual degradation of metrics over time.}
    \label{fig:overtime}
\end{figure}

\noindent \textbf{Per-window evaluation.} After this, we next evaluated how LMDMs perform on such global metrics as a function of time, generating up to 2 minutes of content. Each metric is calculated on a local sliding window of audio context (see App.~\ref{app:evals} for more details). In Fig.~\ref{fig:overtime}, the benefits of ARC-Forcing are on display: without it, in both setups (audio primed and un-primed) nearly every metric gradually degrades as a function of time, while ARC-Forcing significantly mitigates error accumulation. This is true for both Enc-Dec and Block-Causal models on every metric besides FD in non-primed for Block-Causal. We are unsure of the reason as to Block-Causal LMDM's poor performance at higher step count on this metric, and leave further investigation for future work. As Enc-Dec models outperform Block-Causal models in this setting, we focus on this model class for the rest of this work.

\begin{figure}
    \centering
    \includegraphics[width=0.9\linewidth]{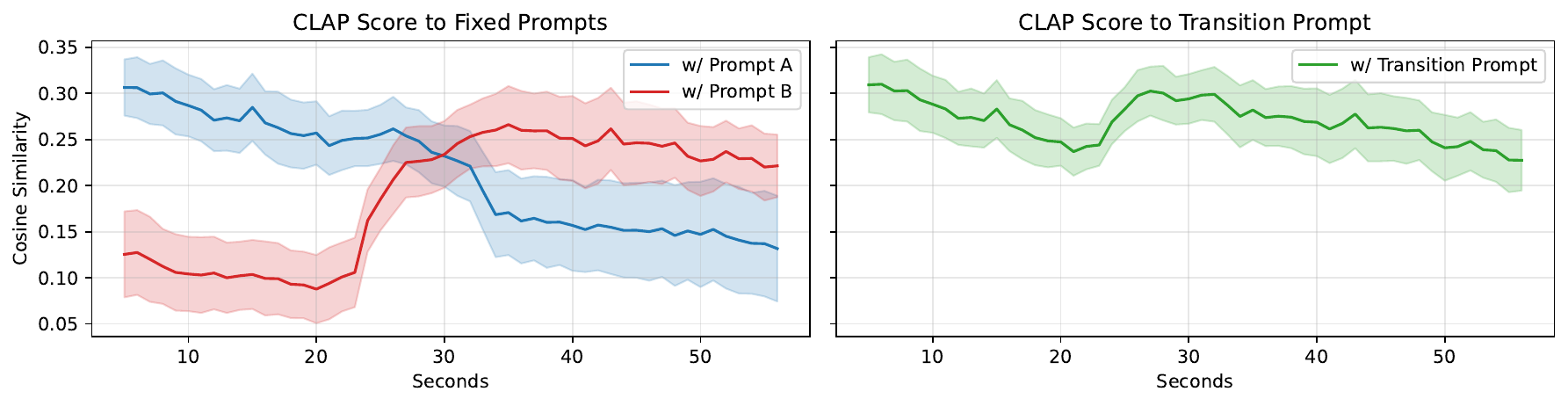}
    \caption{Prompt Transitions using Enc-Dec LMDMs.}
    \label{fig:transition}
\end{figure}

\noindent \textbf{Prompt transition evaluation.} Following~\citet{team2025live}, we then test the ability of LMDMs to perform prompt transitions, where one prompt is cross-faded with another across time. Initially, we found that standard ARC-Forced sampling (which omits CFG) was not enough to break out of the strong conditioning from past generations, and normal CFG resulted in severe over-saturation artifacts at the low step count used. To remedy this, we found two solutions: (1) whenever one prompt crosses a dominant weight over the other prompt, we drop out the first $d=180$ frames of context, reducing the signal of past audio, and (2) we adapt CFG++ \citep{chung2024cfgpp} framework to the standard distilled ``ping-pong'' sampler. This later allowing for increasing text importance without flying off-manifold (see App.~\ref{app:pppp}). In Fig~\ref{fig:transition}, we find that with these modifications LMDMs are able to perform prompt-transition in similar fashion to LMMs.

\subsection{Accompaniment LMDMs}
\begin{wrapfigure}{R}{0.4\textwidth}
    \centering
    \vspace{-1.2em}
    \includegraphics[width=\linewidth]{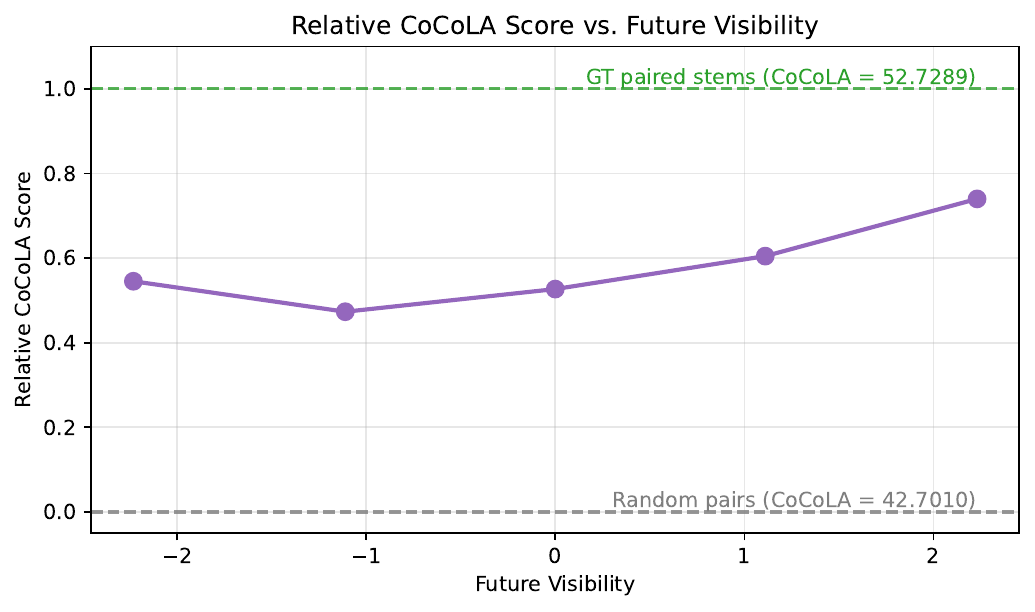}
    \vspace{-2em}
    \caption{Relative CoCoLA score for Accompaniment LMDMs at variable $t_f$.}
    \vspace{-1em}
    \label{fig:cocola}
\end{wrapfigure}
Given our best Enc-Dec setting from the text-conditioned experiments, we next test LMDMs for stem-conditioned accompaniment. 
In this accompaniment-like case, we mainly investigate how LMDMs perform with different amounts of future visibility: in settings where $t_f \ge 0$, the model can condition on stem audio that directly corresponds the the target block but leaves no room for generative latency to stream without a de-syncing from the stem conditions, while $t_f < 0$ settings reduce the amount of stem context the model can see to account for modeling latency and real-time interaction. Here we are primarily concerned with the CoCoLA score \citep{ciranni2025cocola}, which measures inter-stem alignment similar to contrastive methods like CLAP. In Fig.~\ref{fig:cocola}, we show the relative CoCoLA score (normalized between the average score for ground-truth pairs from the same track and fully random pairings) for Enc-Dec ARC-Forced LMDMs with $t_f$ ranging from approximately -2 to 2 seconds. Here we find that while reducing $t_f$ predictably reduces alignment, the model does not collapse to close to random in the $t_f < 0$ case (in contrast to the models from \citep{wu2025streaming}), showing how ARC-Forcing effectively can help mitigate the lack of signal at each block from the stem.

\subsection{Sketch-Conditioned LMDMs}
We leverage the Enc-Dec setting to trained LMDMs on Jamendo with sketch like controls with details in App.~\ref{app:sketch}. 
We evaluate the model on $\approx$11s rollouts with sketch controls extracted from Vocal, Bass, Drums, and other stems from the MusDB \citep{musdb18} test set ($N = 200$) and shuffled captions from MusicCaps \citep{agostinelli2023musiclm}; all generations use no audio prefixes. We choose this rollout strategy and dataset to mirror a real time live interaction in line with our demonstrated use case in Sec.~\ref{sec:userstudy}. In all, we observe comparable control following with respect to an offline bidirectional model we trained with manageable performance degradation for smaller block-sizes. 

\begin{table}[t]
\centering
\footnotesize
\setlength{\tabcolsep}{4pt}
\begin{tabular}{lcccc|ccc|ccc}
\toprule
\textbf{Method} & \textbf{D-NFE} & \textbf{Blocks} & \textbf{Sampler} & \textbf{+AF?} & \textbf{FD}$\downarrow$ &\textbf{KL}$\downarrow$ & \textbf{CLAP}$\uparrow$ & \textbf{Mel}$\uparrow$ &\textbf{Rhy}$\uparrow$ & \textbf{Dyn}$\uparrow$ \\
\midrule
LMDM (ED) & 50 & 5 & Euler & \xmark & 101.01 & 1.52 & \textbf{0.23} & 0.26 & 0.45 & 0.46 \\
LMDM (ED) & 8 & 5 & Ping-Pong & \cmark & 181.79 & 1.24 & 0.14 & 0.27 & 0.45 & 0.45 \\
LMDM (ED-U230) & 50 & 24 & Euler & \xmark & 126.41 & 1.70 & 0.23 & 0.18 & 0.42 & 0.28 \\
LMDM (ED-U230) & 8 & 24 & Ping-Pong & \cmark & 162.38 & 1.32 & 0.15 & 0.21 & 0.42 & 0.38 \\
\midrule
(Bidir) Flow Model & 50 & 1 & Euler & \xmark & \textbf{78.51} & \textbf{1.23} & 0.19 & \textbf{0.33} & \textbf{0.48} & \textbf{0.57} \\
\bottomrule
\end{tabular}
\caption{Distributional and control-following metrics on MUSDB18 test stems (ED and Bidir variants) for the sketch conditioned models with Control evaluation metrics from \cite{tsai2025musecontrollite}. }
\label{tab:jamendo_distributional}
\vspace{-3em}
\end{table}

\subsection{Live Musician Interaction with Sketch-Conditioned LMDMs}\label{sec:userstudy}

We next deploy sketch-conditioned LMDMs as a real-time generative delay effect: the system takes the last block of a musician's audio, computes sketch controls, and schedules the generated block at a fixed offset into the future. As our goal is to move away from high-resource settings and put LMDMs in the hands of real musician, we manage low-resource on-device latency by exporting models via ONNX and embedding forward passes in a custom C++/JUCE application (\ref{fig:interface}).

\noindent \textbf{Functionality.} The LMDM operates as a block-processing audio effect with block size $S$ seconds: it buffers $S$ seconds of input, computes sketch features, runs inference in $\tau_\theta$ seconds, and writes a $S$-second output block, inducing a fixed delay of $\Delta = S + \tau_\theta$ where $\tau_\theta < S$ must hold for gapless playback. Using a 10-latent sketch-based LMDM trained on MTG-Jamendo~\citep{bogdanov2019mtg}, we achieve $\Delta < 1$ second by distilling to 8 steps through ARC-Forcing and deploying with ONNX-exported DiT and VAE in our C++/JUCE app.

\noindent \textbf{Participants and Setup.} We recruited three instrumentalists from a fellowship program at our institution: a saxophonist (P1), a guitarist (P2), and a cellist (P3). P1 and P2 each tried the self-forced Jamendo LMDM as a generative delay ($\sim$1\,s delay) and a foley-like LMDM finetuned on FSD50k~\citep{fonseca2021fsd50k} ($\sim$3\,s delay), with two conditioning modalities for the Jamendo model: sketch controls from the musician's solo signal, or from a mix of the musician with a drum loop to encourage rhythmic consistency. With P3, we collaborated on a composed performance built around an LMDM finetuned on humpback whale songs~\citep{sayigh2016watkins}, culminating in a public concert. Each session ran approximately one hour and included mini-jams and a closing interview; videos are provided in supplementary material.

\noindent \textbf{Responsiveness and Musical Dialogue.}
A recurring theme was the sense that the LMDM behaves as a musical partner rather than a simple effect. P2 noted that the Jamendo model \textit{``both follows you and accurately throws out new ideas,''} adding that \textit{``even if you stay relatively static, it'll reference your playing while adding something different.''} P3 discovered a similar dialogic quality in the whale model, describing the challenge of deciding \textit{``when to initiate, when to answer, and when to simply play my own line and trust something interesting might follow.''} 
Over time, P3 learned to shape the model's responses by manipulating pitch contour and dynamics.
A highlight was building toward an emotional peak in their own playing and hearing the whale responses follow.

\noindent \textbf{Timbral Exploration.}
The musicians also valued the timbral range LMDMs afforded beyond their instruments. P1 discovered that the Jamendo model would shift from bright synth-like mimicry of their saxophone in upper registers to deep bass tones when he played below a certain range. P3 found the whale model compelling for its \textit{``constrained unpredictability''}: she learned to expect two call types whose pitch neighborhood was similar but rarely exact, and gradually developed strategies for triggering responses that fit each musical moment. The foley models expanded this palette further, as P1 used a ``wind chime'' prompt to complement an improvisation of extended saxophone techniques.

\noindent \textbf{Challenges.} For the Jamendo self-forcing model, musicians noted that text-prompt following quickly regressed toward a generic EDM sound during live use, even for prompts like ``disco'' or ``rock and roll''; as this does not appear in offline evaluation, we suspect it is related to our ONNX pipeline. The foley models followed text prompts well but struggled with CQT control, likely because much of FSD50k lacks strong fundamental frequencies, making the top-$k$ CQT uninformative for this domain. 

\section{Limitations and Discussion}

Many challenges remain in improving LMDMs as both standard streaming generation models and real musical tools. Unsurprisingly, we find that the  text-conditioned and sketch-conditioned LMDMs show a high quality bias towards genres in its training data: given MTG-Jamendo's \citep{bogdanov2019mtg} over-representation of electronic dance music (EDM), our LMDMs trained on such data perform considerably better with EDM-like prompts compared to worse represented genres like country or jazz. Additionally, LMDMs in their current form are broadly more responsive to the underlying past clean content than to the input text features, which motivates future work in increasing the reactivity of the model with respect to injected text conditions. We also broadly find that the output quality of LMDMs still trail large frontier models like Suno, leaving much room for closing the quality gap between fast interactive models and opaque proprietary systems.

This being said, we note that LMDMs, and interactive streaming music models more broadly, offer a growing future orthogonal to the scaling of offline Text2Song systems. As hard latency requirements and accessibility on consumer hardware set a fundamental cap on the capacity of interactive models relative to large offline song generators, striving for parity with such systems may not only be a lost cause but actively contradictory to their use for musicians. As we aim for true generative instruments, the unique ways in which models like LMDMs can succeed and fail offer the capacity for ``creative misuse'' \citep{tokui2025surfing} (e.g.~circuit bending, tuned 808 kicks), letting musicians play with such systems and develop their own modes of interaction \emph{unintended} by model builders. The space of live music systems is continuing to evolve into its own unique discipline \citep{kim2026design}, and we are hopeful that such work will increasingly center musicians and their interaction as the core driver of innovation.

\section{Conclusion}

We introduce Live Music Diffusion Models, showing that open-source audio diffusion models can be repurposed into interactive streaming generators through simple routing and attention masking modifications that enable KV-Caching over both diffusion steps and time. Our ARC-Forcing post-training recipe provides stable, RL-free global supervision on multi-block rollouts, significantly mitigating error accumulation. LMDMs achieve competitive quality with drastically reduced latency, running on consumer hardware at a fraction of the parameter count and training cost of discrete-AR alternatives. By deploying LMDMs as a generative delay with real musicians, we demonstrated that these models can serve not just as generators but as responsive musical partners, opening a new design space at the intersection of generative AI and live performance. However, our musician studies consistently highlighted that further latency reduction could encourage more flexible interactions, and achieving sub-second block sizes will likely require advances in causal audio codecs and architectural efficiency beyond the modifications presented here.

\newpage

\bibliography{example_paper}
\bibliographystyle{neurips_2026}

\newpage 

\appendix

\section*{Contributions and Acknowledgments}

\textbf{Zachary Novack} -- Project Lead, Algorithmic Methodology, Text- and Stem-Conditioned Model Development

\textbf{Stephen Brade} -- Project Co-Lead, Sketch-Conditioned Model Development, On-Device Model Wrangling,  Live API Development, Artist Collaboration

\textbf{Haven Kim} -- Data collection and Pre-processing, Evaluation Design and Development, Project Ideation

\textbf{Hugo Flores García} -- Lead Live API Creation and Development, Project Ideation, Artist Collaboration

\textbf{Nithya Shikarpur} -- Live API Development, Artist Collaboration, Project Ideation

\textbf{Chinmay Talegaonkar} -- Algorithmic Methodology, Project Ideation

\textbf{Suwan Kim} -- Live API Development, Artist Collaboration

\textbf{Valerie K. Chen} -- Artist Collaboration

\textbf{Julian McAuley} -- Project Support and Advising

\textbf{Taylor Berg-Kirkpatrick} -- Project Support and Advising

\textbf{Cheng-Zhi Anna Huang} -- Project Support and Advising

We'd like to thank Petros Karypis for their help on debugging a pesky RoPE implementation, Shih-Lun Wu for paper feedback, and Sebastian Franjou and Matthew Michalek for helping demo our sketch-conditioned system.   

\section{Experimental and Evaluation Protocol}\label{app:evals}

\subsection{Evaluation Metrics}

Following prior works~\citep{evans2024open, team2025live}, we report three metrics: Fr\'echet Distance over OpenL3 embeddings (FD-OpenL3)~\citep{cramer2019openl3}, KL divergence over PaSST logits (KL-PaSST)~\citep{koutini2021efficient}, which jointly measure quality and distributional fit, and CLAP score~\citep{wu2023large}, which measures audio--text similarity. All three metrics are computed using the toolkit released with~\cite{saito2025soundreactorframelevelonlinevideotoaudio}. Following prior works~\citep{evans2024open, team2025live}, for text-conditioned LMDMs all reference and conditioning come from the Song Describer Dataset (SDD)~\citep{manco2023song}. For latency, we report the number of function evaluations for the decoding process of a single chunk (D-NFE), as well as the Time to First Frame (TTFF), i.e., the wall-clock time until the first audio is output, calculated on an NVIDIA 6000 Pro Blackwell GPU. For accompaniment generation, we use the CoCoLA score \citep{ciranni2025cocola}, which measures inter-stem similarity. For offline sketch-conditioned evaluations, we use the control evaluation suite from \citet{tsai2025musecontrollite}, and use the MusDB \citep{musdb18} as the reference and control source, with captions from MusicCaps \citep{agostinelli2023musiclm}.

\subsection{Training and Inference Setup}

All models are finetuned from the base version of Stable Audio Open Small \citep{Novack2025Fast}, a 340M parameter DiT originally trained on $\approx12$s of latent audio from Freesound. 

\subsubsection{Text-Conditioned Generation}

LMDMs are trained on the MTG-Jamendo dataset \citep{bogdanov2019mtg}, excluding samples from Song Describer. We train all variants on a fixed length of 240 latent frames with a target generation block size of 48. Models are first finetuned from SAO-Small with the context routing and attention mask for 10k iterations with a batch size of 256, taking approximately 8 GPU hours. ARC-Forcing then proceeds for 18k iterations with a batch size of 80. During ARC-Forcing, we perform 12 block rollouts from the model. In the initial finetuning phase, we set $p_{\text{uncond}}$ and $p_{\text{partial}}$ to 0.2 and 0.3 respectively, while in ARC-Forcing we shift them to 0.5 and 0.1 respectively to improve non-primed generation. For the ARC-Forcing discriminator, we finetune SAO-Small on 768 sequence lengths for 10k steps. When ARC-Forced, LMDMs do not use CFG unless otherwise stated, while non-ARC-Forced LMDMs use a CFG of 7.
We report results for two inference settings on 47\,s clips: \emph{audio-primed generation}, where the model is given a caption and the first $s$ frames of the corresponding ground-truth track as a prefix, and \emph{text-only generation}, where only the caption is provided. We refer to these settings as \emph{primed} and \emph{text-only} hereafter.

To observe drift over time, each generation is produced at the same length as its corresponding ground-truth track, and the same three metrics are recomputed inside sliding windows whose size is determined by each backbone's receptive field: FD-OpenL3 with a 1\,s window and 1\,s hop (matching OpenL3's 1\,s training clip), and KL-PaSST and CLAP score with a 10\,s window and 1\,s hop (matching PaSST's \texttt{max\_model\_window} and CLAP's input length). We use 8 sampling steps for both ARC-Forced models in this experiment. 

We compare against the SOTA Magenta-RealTime, as well as Stable Audio Open \citep{evans2024open} and MusicGen-Large \citep{copet2023simple}. We sweep the number of inference steps for ARC-Forced LMDMs exponentially from 1 to 8 (i.e. 1,2,4,8).

Following Live Music Models~\citep{team2025live}, we evaluate prompt transitions on 128 pairs. Because the pairs from \cite{team2025live} are short genre/instrument tags (e.g., \textit{Accordion} $\rightarrow$ \textit{Ambient}) that do not match the caption-style conditioning distribution our models were trained on, we use pairs drawn from 256 prompts at random from SDD~\citep{manco2023song}. The full list of 256 prompts is provided in Appendix~\ref{app:prompt-transition-pairs}. Note here we use CFG++ \citep{chung2024cfgpp} with a weight of 0.7.

\subsubsection{Accompaniment Generation}

Following \citet{wu2025streaming}, we finetune and post-train Enc-Dec LMDMs on the Slakh MIDI dataset of synthesized stems \citep{manilow2019cutting}, where stems from the same piece are randomly sampled as context and target. In this setup, ARC-Forcing only occurs for 8k steps due to observed faster convergence. We consider 5 models at varying future visibilities in intervals of 24 latents (roughly 1.1s) from 2.2 to -2.2. All other hyperparameters match the text-conditioned case. In this setup, we replace the text conditions with the midi program name for the target stem (e.g.~``electric bass'').

\subsubsection{Sketch-Conditioned Generation}\label{app:sketch}

In Tab.~\ref{tab:sketchapp}, we display the models trained for the sketch-conditioned generation task, including the models used for offline evaluation as well as ones from our user study and performance.

\begin{table}[ht]\label{tab:sketchapp}
\small
\centering
\caption{Overview of sketch-based encoder-decoder models trained across datasets and configurations. Datasets: FSD50k~\citep{fonseca2021fsd50k}; $\approx$\,48 minutes of humpback whale song~\citep{sayigh2016watkins}; MTG-Jamendo~\citep{bogdanov2019mtg}}
\label{tab:trained_models}
\begin{tabular}{l c c c c c c}
\toprule
\textbf{Method} & \textbf{Architecture} & \textbf{Dataset} & \textbf{Block Size} & \textbf{+AF?} & \textbf{Eff.\ BS} & \textbf{Steps} \\
\midrule
LMDM (ED) & Enc-Dec & FSD50k            & 208/47 & \xmark & 128 & 120k \\
LMDM (ED) & Enc-Dec & Humpback whale   & 208/47 & \xmark & 128 & 10k \\
\midrule
LMDM (ED) & Enc-Dec & Jamendo             & 192/48 & \xmark          & 128 & 130k \\
LMDM (ED) & Enc-Dec & Jamendo             & 192/48 & \cmark & 288 & 4.3k \\
LMDM (ED) & Enc-Dec & Jamendo             & 230/10 & \xmark         & 128 & 140k \\
LMDM (ED) & Enc-Dec & Jamendo             & 230/10 & \cmark & 288 & 3.0k \\
LMDM (Bidir) & Bidirectional & Jamendo    & 240/-- & \xmark         & 128 & 120k \\
\bottomrule
\end{tabular}
\end{table}

\section{Derivation of Ping-Pong++ (P4) Solver}\label{app:pppp}

Consider the standard stochastic solver used by few-step ``consistency-style'' \cite{song2023consistency, Novack2025Presto, Novack2025Fast} (i.e. trained to output $\mathbf{x}^{(0)}$ from any $\mathbf{x}^{(k)}$, as opposed to arbitrary $\mathbf{x}^{(s)}$ as in \cite{Kim2023ConsistencyTM}), often called the ping-pong sampler. It's update rule (in flow-matching notation) can be written as:

\begin{equation}
\mathbf{x}^{(k_{i-1})} = (1 - k_{i-1}){\mathbf{x}}_{\bm\theta}^w(\mathbf{x}^{(k_i)}, k_i, \mathbf{c}) +k_{i-1} \bm\varepsilon, \quad \bm\varepsilon \sim \mathcal{N}(0, \bm{I}),
\end{equation}

where ${\mathbf{x}}_{\bm\theta}^w(\mathbf{x}^{(k_i)}, k_i, \mathbf{c}) = \mathbf{x}^{(k_i)} - k_i {\mathbf{v}}_{\bm\theta}^w(\mathbf{x}^{(k_i)}, k_i, \mathbf{c})$ (i.e. we stay in ``v-prediction'' following \cite{Novack2025Fast} to ease convergence in adversarial post-training), and ${\mathbf{v}}_{\bm\theta}^w(\mathbf{x}^{(k_i)}, k_i, \mathbf{c}) = {\mathbf{v}}_{\bm\theta}(\mathbf{x}^{(k_i)}, k_i, \varnothing) + w ({\mathbf{v}}_{\bm\theta}(\mathbf{x}^{(k_i)}, k_i, \mathbf{c}) - {\mathbf{v}}_{\bm\theta}(\mathbf{x}^{(k_i)}, k_i, \varnothing))$ for some guidance weight $w > 1$.

The core of CFG++ \cite{chung2024cfgpp} is to formulate sampling where the \emph{denoising} process maximizes adherence to the text prompt, while keeping the \emph{renoising} process unconditional. We can reformulate our pingpong sampler into this denoising-renoising framework as:

\begin{equation}\label{eq:modpp}
\mathbf{x}^{(k_{i-1})} = \underbrace{{\mathbf{x}}_{\bm\theta}^w(\mathbf{x}^{(k_i)}, k_i, \mathbf{c})}_{\text{denoising}} + \underbrace{k_{i-1}(\bm\varepsilon - {\mathbf{x}}_{\bm\theta}^w(\mathbf{x}^{(k_i)}, k_i, \mathbf{c}))}_{\text{renoising}}
\end{equation}

 To work the ping-pong sampler into a CFG++-style sampler, we modify Eq.~\ref{eq:modpp} to use the \emph{unconditional} velocity for the renoising process, forming:

 \begin{equation}
\mathbf{x}^{(k_{i-1})} =      {\mathbf{x}}_{\bm\theta}^\lambda(\mathbf{x}^{(k_i)}, k_i, \mathbf{c}) + k_{i-1}(\bm\varepsilon - {\mathbf{x}}_{\bm\theta}(\mathbf{x}^{(k_i)}, k_i, \varnothing)),
 \end{equation}

where ${\mathbf{x}}_{\bm\theta}(\mathbf{x}^{(k_i)}, k_i, \varnothing) = \mathbf{x}^{(k_i)} - k_i {\mathbf{v}}_{\bm\theta}(\mathbf{x}^{(k_i)}, k_i, \varnothing)$, and $\lambda \in [0, 1]$. This P4 sampler is able to tune inference-time text strength much more stably than with the standard ping-pong sampler with normal CFG, as the standard implicit $k_{i-1}\mathbf{x}^w_{\bm\theta}(x^{(k_i)}, k_i, \mathbf{c})$ causes noticeable sonic artifacts in the few step regime.

\section{Interface Design}

\begin{figure}[ht]
    \centering
    \includegraphics[width=\linewidth]{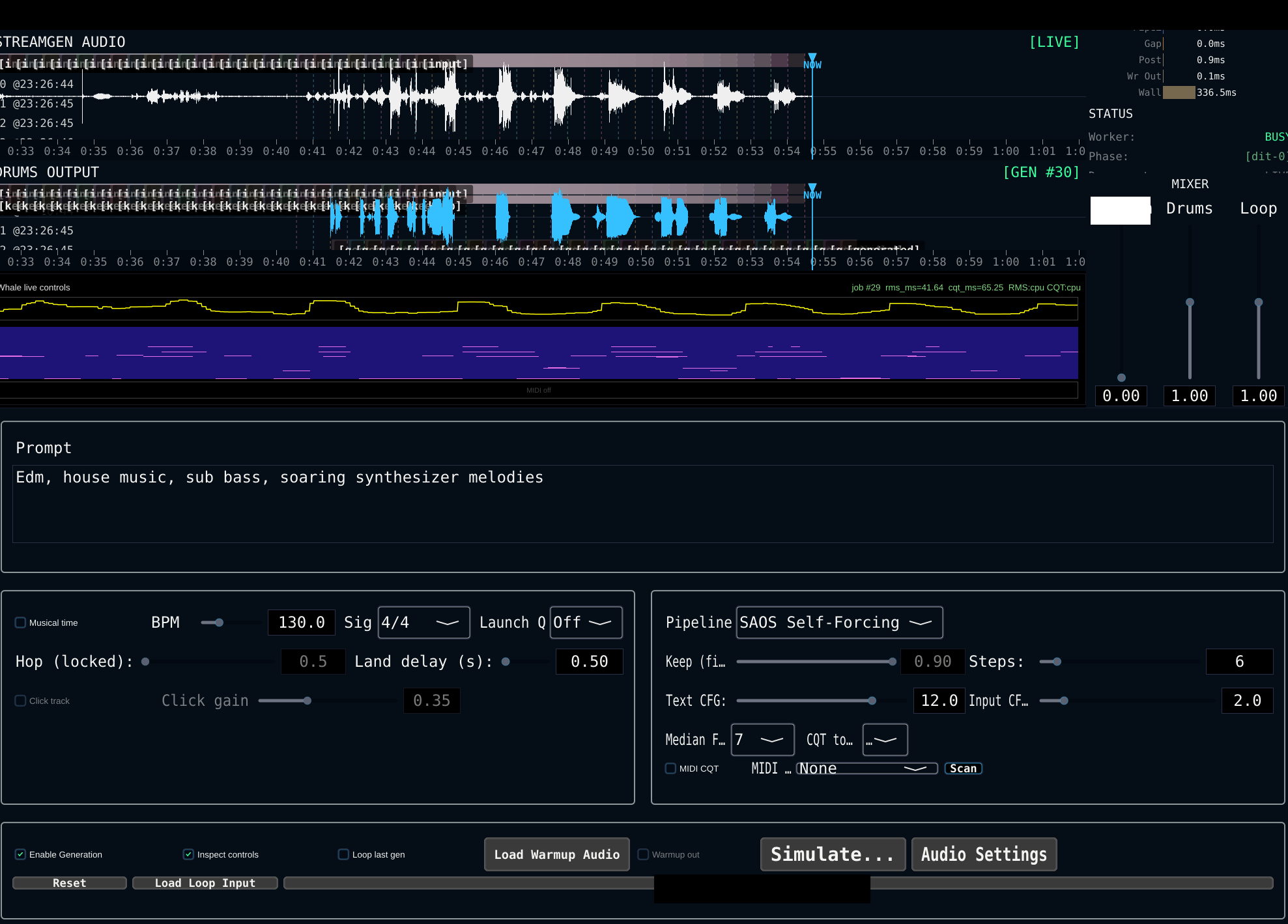}
    \caption{User interface of the system built in JUCE leveraged in the user studies and performances.}
    \label{fig:interface}
\end{figure}

As shown in Figure~\ref{fig:interface}, the interface provides the main interaction components of the application.

\section{Prompt Transition Pairs}\label{app:prompt-transition-pairs}

Table~\ref{tab:prompt-transition-pairs} lists the 128 prompt pairs $(A, B)$ used for the prompt transition evaluation in Section~\ref{sec:experiments}. Both endpoints are drawn from the Song Describer Dataset (SDD)~\cite{manco2023song} captions, sampled from a pool of 256 distinct captions.

\begin{small}
\begin{longtable}{@{}r p{0.45\linewidth} p{0.45\linewidth}@{}}
\caption{Prompt transition pairs used in the cross-prompt continuity evaluation. Each row contains a source prompt $A$ and a target prompt $B$.}
\label{tab:prompt-transition-pairs} \\
\toprule
\# & Prompt $A$ & Prompt $B$ \\
\midrule
\endfirsthead
\multicolumn{3}{c}{\tablename\ \thetable\ -- \textit{Continued from previous page}} \\
\toprule
\# & Prompt $A$ & Prompt $B$ \\
\midrule
\endhead
\midrule \multicolumn{3}{r}{\textit{Continued on next page}} \\
\endfoot
\bottomrule
\endlastfoot
1 & Driving, energetic and positive rock song (male voice) perfect for sport or action. & An alternative rock piece with piano base, drums and male vocal which often does falsettos. \\
2 & Haunting expansive sound as if you are in space & Joyful Christmas song or children's song featuring a bell melody. \\
3 & Bright synth swooshes in the opening followed by a prominent acoustic guitar and a happy-go-lucky bass-line and beat creating a light and breezy feeling evoking a shopping mall or tourist cafe. & Fast-paced heartsick synthwave techno with crisp melodies and repetitive drums \\
4 & Pop rock song with male lead singer containing slight dissonant passages between the lead guitar and the supporting piano chords. & Emotional and intimate lovely French song featuring acoustic guitars and soft male vocals \\
5 & The song as a catchy brass riff and percussions and an upbeat guitar in the background, with a Spanish lyric sung by two singers, one male and one female singers. & The relaxed melody and slow tempo make this song a combination of romantic and peaceful piece \\
6 & Electronic rock high-energy song with vocals but no reverb that draws you in with instrumentation changes, syncopation, and panning. & A funky rock song so high paced and dynamic that it makes one dizzy. \\
7 & Heavy metal song with folk influences, drum and distorted guitars are sustained by a French speaking male voice and a synth lead, in some point the music stops and there is a rain sample. & Calming and tense classical music played only on piano with strong variance in change of dynamics (velocity) \\
8 & A strings orchestra and piano combine in this waltz to give a fantasy feeling. & Country song with acoustic guitar and singing along with slide guitar for embellishments that can be listened to while relaxing at home or just driving \\
9 & Creepy electronic track gives a sensation of suspense and intrigue. & A romantic song which features a couple who just reconciled after going through individual hardships or challenge in their relationship \\
10 & A typical punk rock song of the 2000s, sung by a male voice with a positive and energetic attitude. & A peaceful piano piece for relaxing by the fireplace. \\
11 & A slow swingy song with a male vocal, bluesy clean guitar and smooth drums. & It is an instrumental piano song, with a moody relaxing and gentle tone \\
12 & Instrumental piece with some baroque-era musical arrangements that could well be used in the opening credits of a period film. & Serene, but slightly tense piano piece played at a moderate tempo \\
13 & an uplifting jazz song that makes your head shake & Psychedelic rock with raggae rhythm featuring electric guitar with phaser effects, bass and drums but no vocals \\
14 & A warm and slow paced song that has an R\&B base invites to get closer. & Cheerful French love song in a reggae style, and with chorus in English. \\
15 & Piano ternary piece with a repetitive and dynamic pattern that modulate overtime with some 3-for-2 parts & Its an ambient song, with electronic elements like a synth, delayed piano, sine wave oscillators, no vocal, can come under the background, mysterious vibe, relaxing as well \\
16 & Electropop track with lots of synths and a female singer with an eastern european accent, making you want to dance at the club. & Neo-soul song with nylon string guitar and female vocals in French \\
17 & Mid tempo blues swing song that one might find in a casual bar setting for people to dance to & This type of electronic music can be used in a dance club since it builds up fairly quickly and has a dance beat \\
18 & Ominous 2010s hip-hop beat with FM-style bass and discordant sound design and non-english rapping. & Nervous gypsy instrumental song with jazzy acoustic guitars. \\
19 & Electronic instrumental that has a consistent beat and a melody that at times has a descending pattern which can be used for a soundtrack & Slow tempo pop song using only piano and vocals with sad lyrics \\
20 & Progressive electronic song with an intro of African percussions. & Positive instrumental pop song with a strong rhythm and brass section. \\
21 & Upbeat rock guitar song with a punk feel about love & Trance experimental electronic track with a weird sounding vocal sampled lead, hollow drums and a 8-bit sounding beat. \\
22 & This son is pumping with heavy percussion layers and an energetic female vocal. & This is an electronic track with classic snare-roll build up, four-on-the-floor drums and syncopated synth melodies \\
23 & Groovy instrumental funk rock track with occasional guitar solos that give you a feeling of longing. & A pop-rock love song from the 80's that conveys a quiet sense of positivity. \\
24 & Country drinking song with a ragtime or blusey vibe, featuring a female voice. & Electronic music with a sci fi feel to it like an event is about to start which is mostly based on various percussion as compared to a melody \\
25 & This track starts with some effect making it look old and vintage, before the effect is removed and the singer starts singing louder, supported by drums, bass and electric guitar & This energetic rock song starts with a drumstick countdown and has a catchy guitar riff and male vocalist. \\
26 & Calming instrumental played on bass violin and flute that can be put in the background while doing some work or study & A catchy electropop track featuring a male vocalist with a unique, drawling delivery style and occasional autotune that adds a cool, quirky edge to the energetic, synthesized soundscape, while the catchy pop chorus is sung by a female vocalist. \\
27 & A guitar folk song with husky male voice & Gentle and dreamy love song with acoustic guitars, strings and male vocals. \\
28 & This is an experimental electronic songs, with noisy synths and a male voice speaking a slavic language. & A sweet and fun track featuring slide guitar and a male vocal. \\
29 & A country tune with a male lead voice and some backing vocals, played with several guitars, some of them using slide. & A harsh instrumental rock song which sounds a bit artificial and dull. \\
30 & French folk, singer-song writer piece featuring guitar and voice. & Ambient instrumental piece with a sense of birth from nothing. \\
31 & Mysterious piece of jazz music for a typical film noir scene, played on piano, bass (and with a trumpet solo section). & Country-tinged, piano driven pop with a female lead singer \\
32 & This is a calm swing music with drums, synth and trumpet responding to each other and makes you want to bounce. & Quivering male voice over solo acoustic guitar, indie folk rock, playful. \\
33 & initially calm and slightly solemn solo piano with pronounced quick double attacks on some notes making it more lively later on & Sounds like a vaporwave song with 80s-style synths, a steady, driving beat with hand percussion and nature sounds in the background. \\
34 & Slow tempo pop song using only vocals and piano with lyrics around love & Progressive rock song with a driving synth base and slurred high pitched male voice. \\
35 & Elegant and pure sounds that convey a sense of order and cleanliness & A lively and happy ska song featuring energetic brass and children vocals. \\
36 & This song starts with an ambient pad intro with a hip-hop influenced beat dropping halfway through. & This is a very fast and dynamic gipsy track with trumpets and voices, that makes you want to jump. \\
37 & Filmic uplifiting non-vocal orchestral piece which builds into a synth and guitar based track & Rock song with a heavy guitar and bass riff played mostly on down strokes with an energizing beat supplemented with vocals \\
38 & An instrumental Christmas carol with a trumpet that leads the melody and a choir of "ahhs". & Pop song that is initially sung by a male voice and then a female voice which carries most of the melody that has an upbeat tempo but sad style \\
39 & An alternative rock song with a male lead singer,  drums and an electric guitar with a little distorsion. & A twisty nice melody song by a slide electric guitar on top of acoustic chords later accompanied with a ukelele. \\
40 & A nasal synth line opens the song before a choir of women singers give an emphatic lyrical performance about antagonism and hurt; a deep synthetic bass runs through the track and the drums get more active as the song continues. & Christmas carol with only instruments feels like in a Disney land fill with toys. \\
41 & A soothing track with a mellow synth sound enriched progressively by an upbeat drum machine and robot-like vocalizations. & energetic ska track with driving guitars and drums and brass featuring raw and scratchy vocals \\
42 & this song starts with an emotional piano melody and then drops into a kind of DnB track with drum break samples and a thumping bass, keeping the piano and later on it has a drum break, whe a violin is added. & Classic video instruction background song with major chord and ukelele. \\
43 & An epic soundtrack with drums and choirs that conveys a sense of tension, defiance and danger. & A happy Latin song with a strong rhythmic and percussive component that invites you to dance and enjoy. \\
44 & Dancefloor, edm influenced instrumental, driven by a piano house progression layered with trance saws & This instrumental song is so calm that makes me feel sleepy \\
45 & A smooth blues song gets exciting as the vocal joins and lets us through his meandering melody. & Energetic bluesy song with a harmonica and horn section in musical dialogue. \\
46 & A loud rock/metal song in French with guitars and drums & Instrumental with mostly piano that has a melancholic feel which can be heard alone \\
47 & An ambient track perfect for meditations and focus tasks. & A fun song with guitar, drums, brass instruments and a male vocal. \\
48 & Instrumental rock song that begins with an ambient intro and then progresses through various sections of electric guitars that intensify over time. & Groovy rhythm with the guitar especially when the bass kicks in, un-noticeable absence of loud snare and kick, poppy and catchy \\
49 & Elegant and fragile piece of orchestral instrumental music for a film soundtrack set in the Middle Ages or in an Asian country, with a melody played by a violin accompanied by plucked strings. & This is a experimental piece or sound effect with quiet noises of running water and someone whispering. \\
50 & Upbeat acoustic guitar song with country style singing and vocals along with harmonica plus whistling at the end & Rock style track with simple bassline, heavy guitars, and a male rock vocalist. \\
51 & A traditional heavy metal riff intro with a transition to grunge-like verse. & 90s hip hop with a moody synth which gives a very ominous but danceable vibe to it. \\
52 & Classical symphony, sounding like someone is doing something hastily & A blues song with crisp modern production, featuring punchy drums, lots of slide guitar and a passionate male singer. \\
53 & Jungle or forest fauna sounds followed by floaty ethnic flute solo on top of some hopeful synthesizer harmonies. & Adventurous and curiosity inspiring soundtrack consisting of bells, flutes, strings, and choir. \\
54 & Smooth chill electronic music featuring calming flute and techno beats. & Fast tempo percussion with an energetic beat and a vocal melody \\
55 & Childish and innocent orchestral piece of music that conveys joy and happiness and gives a sense of beginning. & Upbeat song that has a humming riff accompanied by guitar licks that can be used for a casual listening setting \\
56 & Happy optimistic instrumental song with whistles, xylophone and ukulele that can be used to clap and dance along to & Strummed acoustic guitar and female vocal on an uplifting pop song performed by real musicians. \\
57 & A folk song with a bitter-sweet acoustic guitar and a pleasant sounding male vocalist, that turns into a melancholic duet with a female voice. & A mellow and joyful piece of classical music for solo piano \\
58 & A Latin American song with accordion and trumpets and a swaying, but heavy feel. & This is pop song with drums, synth, electirc guitar and some lyrics, with a classical chord progression and a memorable melody. \\
59 & Instrumental mainly focused on clean electric guitar sound riffs and melody which might be fun to listen to on the background & Classical piano piece in major key with a slow start, played beautifully without other instruments, making you close your eyes and find inner peace. \\
60 & A medium-paced EDM-style song with an intro by a synthesizer, features some ambient sound and simple drum. & A gentle folk song with an acoustic guitar and a male voice that conveys a sense of fragility and hope. \\
61 & Slow-paced guitar ballad with clean guitar solos interspersed throughout the composition & Indie folk-rock song sung by a young male vocal with things to say. \\
62 & A French folk track backed by acoustic guitar & Guitar heavy folk song with accompanying male vocalist talking with introspection, not a particular happy song \\
63 & A whimsical string arrangement that feels like bouncing through some woods on an adventure in a video game, with a triumphant finish. & Fast tempo electronic music with a melody that repeats consistently \\
64 & Driving rock song with an energetic chorus, featuring heavily distorted guitars and male vocals. & A dreamy and ethereal piece of electronic music with a Spanish guitar. \\
65 & Hip hop track with a subtle reggaeton feel, with a male rapper and a female soul choir in the background. & Innocent and playful classical music for orchestra featuring wind musical instruments such as oboes or fagots. \\
66 & This is an instrumental track based on electronic samples that can be used before the start of a movie to give a doomsday or post apocalypse feel & A folksy ditty featuring a warbling woman narrating a story over a variety of jamming, Western-sounding instruments, which evokes a nostalgic sense of older country, bluesy rock. \\
67 & A positive tune with a surprising mix of instruments: acoustic guitar, tabla and sitar & a rock genre  song with fast tempo guitar riffs and drums with a processing voice singing \\
68 & Pop song with claps, piano chords and vocals & This song has a murky, underwater sound with panicked vocals \\
69 & This smooth jazz track which features a sleepy English vocal is pleasant to listen and perfect for nightdreaming. & depressive music with only guitar and a sad voice for guys using drugs would be ideal for a rainy day to be even more sad \\
70 & A rap song with two male voices and sounds in the background that loosely resemble a siren. & Electro dance song to play in the pub to cheer up the crowd. \\
71 & A very energetic and bright funk rock song which features a noticeably solid guitar licks. & Pop love song that would accompany perfectly a morning run. \\
72 & The song start with a dark guitar riff but quickly turns into French EDM mixed with downtempo, with a strong beat and vocal samples and pads morphing through the song. & French-language song with a jazzy, late-night vibe featuring a male duet. \\
73 & One cannot avoid moving the feet and neck listening to this fast and loopy brazilian tune. & Portuguese-spoken ska track with usual melody and instruments. \\
74 & 1990s techno with cliche minor key progression and tacky synthetic saxophone lead & The song lyrics is in Spanish and has a salsa rhythm that makes you want to have fun and dance \\
75 & Wonderfully emotional soundtrack with a violin carrying the main melody. & A ska song with all the usual suspects except for a slow and kind of melancholic intro. \\
76 & A ballad song with an acoustic guitar and a male voice singing french words. & energetic rock song about love with guitar and piano \\
77 & A wobbly funky track with a foreground bass line and rhythms made with vocal just makes one feel alright. & Grand, ambitious movie music with powerful, march-like drum beats, as well as brass melody. \\
78 & A positive and enthusiastic pop-rock song from the 80s featuring reverbered guitars. & A frantic rock intro with bass and drums turns into a happy and energetic ska punk song with a male vocalist. \\
79 & Piano ballad that could be used in a ballet, accompanied by percussion and a female vocalist with backing harmonies & Introspective, raw rock song with organic acoustic guitars and a raspy voice \\
80 & Lush vocals on well-reverbed guitars give a nice sensation at first but then makes the listener ask for more. & cheerful happy music played on a piano for relaxing \\
81 & Male vocalist with a raspy voice singing over melancholic piano chords and drums increasing in intensity, with a slighty dissonant chorus featuring distorted guitars. & slow and arpeggiated solo piano with a boatload of reverb \\
82 & a sinister medieval-sounding piece with a synthesized flute or organ and strong downbeats at a marching pace. & Instrumental on acoustic and electric guitar with a pleasant feel that can be played in a cafe \\
83 & A pleasant instrumental folk song with acoustic guitars and violins that conveys a sense of peace and happiness. & A thrilling instrumental track featuring a series of stringed instruments being gradually built into the song, culminating in eastern european vibe section. \\
84 & Upbeat hip-hop with trap vibes and bubbly bass synth. & 2000's style urban pop song with synthesised pad and drums. \\
85 & This a dark artsy metal piece featuring plenty of sound effects. & This sentimental rock track features a clean electric guitar with heavy echoes and troubled male vocals. \\
86 & A driving french song featuring fast guitars playing. & 8-bit melody brings one back to the arcade saloons while keeping the desire to dance. \\
87 & A fuzzy and dampened electronic piece which progressively brightens throughout the track. & This track featuring a solo piano gives a sense of safety, determinedness and vulnerability. \\
88 & Upbeat fast tempo with a blues rock feel that one can dance & Acoustic guitar that overtakes the female vocal line) starts the song, as a rimshot-heavy drum groove emerges; warbly vocals bring in the chorus, which is nostalgic and reminiscent. \\
89 & A lively and fun ska song in Spanish with energetic brass sections. & Classical solo piano music, romantic period. Complex harmonic textures, the pièce reminds of a joyful but melancholic moment, with cold weather outside and the bite of winter at the door, sitting around a fireplace in the living room. \\
90 & Energetic pop rock track that will surely keep the dancefloor moving in a live concert. & Upbeat latin/salsa song with Spanish lyrics, accordion, and flamenco  guitar suitable for a warm and hectic dance floor \\
91 & This is an alternative rock song with slow tempo and guitar, male vocals and drums. & Synth-y, spacey art-pop track with a catchy beat \\
92 & It's a folk song with country vibe, major chord, and rock drum kit. & Indie song with a synth effect riff along with vocals \\
93 & A french song that opens with an answering machine sample and then evolves into a alt rock track with frenetic guitar split by a accordion-like synth track & A driving indie rock song sung in Spanish by a male vocalist. \\
94 & This song makes you feel like entering to a haunted mansion or magical village. & A catchy bassline with a drum rhythm sequence introduces the song, followed by a simple chord cadence played on a piano, with some vocals with no lyrics. \\
95 & Electronic disco music about love that builds up quickly and has a dance beat to it & A rock song that transitions between calming humming melody and regular rock singing giving a change of feel throughout the song \\
96 & Male hip-hop track with a tribal vibe featuring african percussion. & Typical energetic and positive EDM song from the early 10s that could be played in a discotheque on a summer night. \\
97 & A fun upbeat drums section playing with a keyboard and some electronic horn type noises, feels like the automatic song on a keyboard that you'd play in music lessons & An harmonised male vocal sample intro, a downtempo electronic beat, a male sung chorus riff followed by a male rapped verse. \\
98 & A pop style synthesized instrumental track with mellow, chorus-like top line, and a prominent drum beat. & A sad pop song using piano and vocals that has lyrics around someone they love \\
99 & Instrumental jazz ensemble with a bossa nova rhythmic vibe that conveys a sense of positivity. & Live recording of a french blues music that starts with a male voice talking alone, and then is followed by a fast-paced blues music with drums, trumpets, harmonica and electric guitar \\
100 & Latin chill song with a bosa nova flavour, accompanied by a melodic trumpet and sung by a male voice. & The rap song has a catchy riff with percussion based on claps and beats but there is a vocal rap on top that goes into a chorus that one can sing along to \\
101 & Electronic music that can be listened to in a passive setting & An instrumental world fusion track with prominent reggae elements. \\
102 & Upbeat commercial-sounding song with a simple drum rhythm, guitar chords and a simple but spacey-sounding synth melody & Futuristic experimental song with piano over a drum and bass rhythm, with an accordion added towards the middle of the song. \\
103 & A ukulele instrumental track sounding like children's song, with a persistent clap percussion and some different riffs on the same theme played in sequence by a glockenspiel, a synth and a slide guitar. & Rap song full of pop elements and sound effects, quite classical and simple dispite off the complex post production \\
104 & Repetitive, sad feeling pop song with a piano intro, finger snapping and female vocal samples. & A musical call to be in the here and now with this deeply spiritual piece. \\
105 & Heavy distorted lead guitar with sparsely supporting piano and distant sounding drums in the background & Melodic and heartfelt piano ballad performed by a female voice accompanied by piano. \\
106 & A whimsy and unserious piano melody, underpinned by consistent snares with occasional off-beat hihats,  with a woman singing in a sometimes childish and sometimes sultry voice & a piano ballad with a female singer evoking nostalgia and disillusioned love \\
107 & Repetitive dancefloor track with loud synths and a simple beat & Gentle folk song with male vocals and arpeggiated acoustic guitar. \\
108 & An romantic song with only piano and vocals that is contemplative yet relaxing & This song starts with a strange french conversation and chain sounds and then gives way to a very heavy and intense synth metal section. \\
109 & EDM fast-paced non-vocal track with light four-on-the-floor beat predominantly based around wobble saw bass line and synth melody & This track is hispanic genre, with spanish lyrics, it has a sort of romantic, love theme, with trumpets, guitar and drums. \\
110 & arythmic piano accompanied by some catchy lyrics sung by an emotionally charged man. & This track starts creating an ambiental atmosphere on which a female singer starts singing, evoking a misterious and sinister character. \\
111 & This track sounds like the backing track to an informative video about a campaign or a new product. & A luminous and moving instrumental piece for piano. \\
112 & This midi electronic instrumental has a hopeful and optimistic vibe with melodies that mostly repeat for 4 bars & This song starts with a piano that can be heard from far but then gets exciting with guitar and a fast-paced male vocal. \\
113 & Rising sweling violin instrumental that elicites calm floating ephemeral emotion. & Rock song starting with guitar line, with a catchy refrain. \\
114 & A dark and slow electronic track makes one think of scary dungeons filled with slimy skulls. & Two electric guitars in conversation with each other, one with a wua-wua effect and the other with a strong delay effect. \\
115 & Warm repeating melody which seems to be a folk music playing by a band where they greets new visitor. & Contemporary trendy optimistic indie pop, with dirty drums, happy guitar comping and synthesizer solo \\
116 & Typical guitar-folk song with no lyrics, but a choir of voices being used for giving texture. & Instrumental song with a sad but intense romantic Latin vibe, the melody is carried by a piano. \\
117 & Orchestral music with a slow and steady pace that sounds like a soundtrack to a nature movie or documentary. & A modern country song with romantic themed lyrics. \\
118 & A track with brit rock essence accompanied with pan flute melody. & Instrumental ambient track with an 80s chillout vibe, featuring bongos and a piano melody. \\
119 & Synthetic orchestral music with hopeful, yearning harmonies, fast pizzicato motifs and woodwind lead & This generic pop/ rock song features a man singing with a nasal voice, with a tired and boring beat behind it. \\
120 & upbeat synthesized guitar pop instrumental track, with consistent tempo and rhythm & A retro-futurist drum machine groove drenched in bubbly synthetic sound effects and a hint of an acid bassline. \\
121 & Instrumental song played by an ensemble of cheerful acoustic guitars, giving the feeling that all is well and nothing bad is going to happen. & Unnerving mix of electronic and rock music that has a gloomy vocal feel and turns into screaming in the chorus \\
122 & Intriguing, slowly progressing electronic tune perfectly match for an indie platform game & Ambient song with duduk and oriental drums perfect for starting a trip to east. \\
123 & Pop guitar strumming with a raspy whispery low female voice & Alternative / experimental rock song with male vocals an a futuristic dreamy vibe. \\
124 & Calm sitar and Indian tabla with dramatic synthetic strings background & A joyful classical track performed on a grand piano. \\
125 & EDM pop song with an energetic an positive mood. & A cinematic piece with a very bright piano, later joined by a drum machine, hand  clapping and a violin synth sound which makes the track sound a little bit more oriental. \\
126 & bluesy guitar with a slow repetitive rythm in a smoky room in latin america & Agitated West Coast rock with brief bass solo and British influence \\
127 & Eurodance pop track, with a simple rough synth chord progression and a straightforward drum beat punctuated by a female voice sample. & This is an energetic and positive rock song with guitars, keyboard, drums and a male vocal. \\
128 & A delightful chord progression on an acoustic guitar later accompanied by an ukelele and harmonica. & Indie rock most likely a 4 piece band that someone can listen to while driving \\
\end{longtable}
\end{small}

\end{document}